\documentclass[11pt]{elsarticle}

\usepackage{amsmath,amsthm} 
\usepackage{amssymb,mathrsfs,bm} 
\usepackage{a4wide} 
\usepackage{graphicx} 
\usepackage{color,subfigure} 
\usepackage{enumerate}  

\DeclareMathAlphabet{\mathpzc}{OT1}{pzc}{m}{it}
\newcommand{\dps}{\displaystyle } 
\newcommand{\rme}{\mathrm{e}}

\newcommand{\eps}{\varepsilon}
\newcommand{\pr}{\parallel}
\renewcommand{\leq}{\leqslant}
\renewcommand{\geq}{\geqslant}
\newcommand{\dt}{{\Delta t}}
\renewcommand{\pr}{\parallel}
\newcommand{\To}{T_*}
\newcommand{\Cv}{C_v^\infty}
\newcommand{\Cvo}{C_v^0}
\newcommand{\cD}{\mathcal{D}}
\newcommand{\cE}{\mathcal{E}}
\newcommand{\kB}{k_\mathrm{B}}
\newcommand{\epsref}{\eps_\mathrm{ref}}
\newcommand{\Tinit}{T_\mathrm{init}}
\newcommand{\betainit}{\beta_\mathrm{init}}
\newcommand{\bgamma}{\bm{\gamma}}
\newcommand{\bsigma}{\bm{\sigma}}
\newcommand{\br}{\bm{r}}
\newcommand{\bB}{\bm{B}}
\newcommand{\bG}{\bm{G}}

\begin{document}

\journal{J. Comput. Phys.}

\begin{frontmatter}



\title{Stable schemes for dissipative particle dynamics with conserved energy}


\author{Gabriel Stoltz}

\address{Universit\'e Paris-Est, CERMICS (ENPC), INRIA, F-77455 Marne-la-Vall\'ee}

\begin{abstract}
This article presents a new numerical scheme for the discretization of dissipative particle dynamics with conserved energy. The key idea is to reduce elementary pairwise stochastic dynamics (either fluctuation/dissipation or thermal conduction) to effective single-variable dynamics, and to approximate the solution of these dynamics with one step of a Metropolis-Hastings algorithm. This ensures by construction that no negative internal energies are encountered during the simulation, and hence allows to increase the admissible timesteps to integrate the dynamics, even for systems with small heat capacities. Stability is only limited by the Hamiltonian part of the dynamics, which suggests resorting to multiple timestep strategies where the stochastic part is integrated less frequently than the Hamiltonian one.
\end{abstract}


\end{frontmatter}

\section{Introduction} 

Dissipative Particle Dynamics (DPD)~\cite{HK92} is a particle-based coarse-grained model in which atoms, molecules or even groups of molecules are represented by a single mesoscale particle. The time evolution of the mesoscale particles is governed by a stochastic differential equation. Dissipative and random forces allow to take into account some effect of the missing degrees of freedom. DPD was put on a firm theoretical ground in~\cite{EW95}. However, it is intrinsically is an equilibrium model, with a prescribed temperature, and cannot be used as such to study nonequilibrium systems. It should be replaced by a dynamics where the fluctuation/dissipation relation is based on variables which evolve in time. DPD with conserved energy (DPDE) is such a model~\cite{AM97,Espanol97}. In the DPDE framework, mesoparticles have an additional degree of freedom, namely an internal energy, which accounts for the energy of the missing degrees of freedom. The dynamics on the internal energies is constructed in order for the total energy of the system to remain constant. DPDE was initially used for thermal transport~\cite{REE98,MBN99}, and later on to simulate shock and detonation waves~\cite{Stoltz06,MSS07,MVDS11}.

While numerous efficient schemes were developed for DPD (see for instance~\cite{LS15} for a review and careful comparison of various schemes), the efficient numerical integration of DPDE still requires some effort. One appealing framework to integrate DPDE, as considered in~\cite{Stoltz06,LBMLM14} for instance, is based on the so-called Shardlow splitting algorithm (SSA) for DPD~\cite{shardlow03}. It consists in decomposing the dynamics into a Hamilonian part and pairwise elementary dynamics - either fluctuation/dissipation or thermal conduction. There is a consensus on the integration of the Hamiltonian part, for which a Verlet scheme~\cite{Verlet} should be used. There is on the other hand no definite way of integrating the fluctuation/dissipation and thermal conduction parts, even when they are split into elementary pairwise dynamics. In particular, to the author's knowledge, for all the numerical schemes currently used, it is observed that negative internal energies may appear when the fluctuation terms are large compared to the heat capacity. This sometimes puts a severe constraint on admissible timesteps. This issue has been explicitly acknowledged by various researchers~\cite{REE98,BM99,MBN99,QH07,HMS16} (and hidden under the rug by others), but no satisfactory answer was found yet. 

Better integration schemes can be obtained by a dedicated treatment of the elementary fluctuation/dissipation and thermal conduction dynamics, instead of resorting to general purpose integration schemes such as Euler--Maruyama. The key observation is that the seemingly $2(d+1)$-dimensional elementary fluctuation/dissipation dynamics can be reduced to an effective one-dimensional dynamics, which can be integrated with a high precision and/or stabilized by a Metropolis--Hastings acceptance/rejection procedure~\cite{MRRTT53,hastings70}. In particular, the Metropolis procedure automatically corrects for negative internal energies. A similar reduction can be performed to obtain an effective one-dimensional dynamics for the elementary pairwise thermal conduction, which is a priori of dimension~2. 

\bigskip

This article is organized as follows. DPDE and the general splitting strategy for its numerical discretization are recalled in Section~\ref{sec:DPDE}. Section~\ref{sec:metropolis} is the core of this work: It is shown there how to numerically integrate elementary pairwise stochastic dynamics in order to exactly sample the invariant measure of DPDE. The resulting numerical method is tested on various systems in Section~\ref{sec:numerics}. Section~\ref{sec:ccl} gathers the conclusions and some perspectives of this work.

\section{Dissipative particle dynamics with conserved energy}
\label{sec:DPDE}

The governing equations of DPDE are recalled in Section~\ref{sec:equations_DPDE}, while Section~\ref{sec:microEOS} discusses microscopic equations of state which allow to model temperature-dependent heat capacities (some technical derivations being postponed to the Appendix~A). A general framework for the numerical integration of DPDE is finally presented in Section~\ref{sec:splitting}.

\subsection{Description of the dynamics}
\label{sec:equations_DPDE}

In dissipative particle dynamics with energy conservation, the variables describing the state of the system are the positions $q=(q_1,\dots,q_N)$ of the $N$ particles, their associated momenta $p=(p_1,\dots,p_N)$ and the corresponding internal energies $\eps = (\eps_1,\dots,\eps_N)$. The positions $q_i$ belong to a position space $\cD$ (typically, a simulation box with periodic boundary conditions), the momenta $p_i$ can assume any value in $\mathbb{R}^d$ (with $d$ the physical dimension), while the internal energies $\eps_i$ are scalar variables which should remain non-negative. Denoting by $V(q)$ the potential energy of the system, the evolution of the variables $(q,p,\eps)$ is governed by the following equations~\cite{AM97,Espanol97}:
\begin{equation}
  \label{eq:DPDE}
  \left\{ \begin{aligned}
    dq_{i} & = \frac{p_{i}}{m_i} \, dt, \\
    dp_{i} & = -\nabla_{q_i} V(q) + \sum_{j \neq i} \left[ -\gamma(\eps_i,\eps_j) \chi^2(r_{ij}) \Big(e_{ij} \cdot v_{ij} \Big) e_{ij} \, dt + \sigma \chi(r_{ij}) e_{ij} \, dW_{ij}\right], \\
    d\eps_{i} & = \sum_{i \neq j} \frac{\chi^2(r_{ij})}{2} \left[ \gamma(\eps_i,\eps_j)\Big(e_{ij} \cdot v_{ij} \Big)^2 - \frac{\sigma^2}{2} \left(\frac{1}{m_i}+\frac{1}{m_j}\right)\right]dt - \frac{\sigma}{2} \chi(r_{ij}) \Big(v_{ij} \cdot e_{ij}\Big) dW_{ij} \\
& \ \ + \sum_{i \neq j}  \kappa \chi^2(r_{ij}) \left(\frac{1}{T_i(\eps_i)}-\frac{1}{T_j(\eps_j)} \right) dt + \sqrt{2\kappa}\chi(r_{ij}) \, d\widetilde{W}_{ij},  
  \end{aligned} \right.
\end{equation}
where $m_i$ is the mass of the $i$th particle, 
\[
e_{ij} = \frac{q_i-q_j}{|q_i-q_j|}
\]
is the unit vector in the direction~$q_i-q_j$, $r_{ij} = |q_i-q_j|$ is the distance between particles~$i$ and~$j$, $\chi$ is a cut-off function, $(W_{ij})_{1 \leq i < j \leq N}$ and $(\widetilde{W}_{ij})_{1 \leq i < j \leq N}$ are two families of independent standard one-dimensional Brownian motions with $W_{ji} = -W_{ij}$ and $\widetilde{W}_{ji} = -\widetilde{W}_{ij}$ for $1 \leq i < j \leq N$. The fluctuation magnitude $\sigma \geq 0$ and the thermal conductivity $\kappa \geq 0$ are fixed (although they could depend on the particle pair). Note that the version of DPDE where the friction forces and fluctuation terms are projected along the lines of center of the dynamics is considered here. The extension of the numerical schemes presented in this work to more general dynamics with both parallel and perpendicular fluctuation/dissipation terms (as in~\cite{JPK08} for DPD) is straightforward; see Appendix~B for precise formulas.

It can be shown that the dynamics preserves the total momentum and the total energy, sum of the mechanical energy~$H$ and of the internal energy:
\[
\cE(q,p,\eps) = H(q,p) + \sum_{i=1}^N \eps_i, \qquad H(q,p) = V(q) + \sum_{i=1}^N \frac{p_i^2}{2m_i},
\]
This is discussed more precisely in Section~\ref{sec:splitting}, where it is shown that the complete DPDE evolution can be separated into elementary dynamics which all preserve the total energy~$\cE$. Moreover, the friction is taken as 
\begin{equation}
\label{eq:FDR1}
  \gamma(\eps_i,\eps_j) = \frac{\sigma^2}{4k_{\rm B}}\left[\frac{1}{T_i(\eps_i)} + \frac{1}{T_j(\eps_j)}\right],
\end{equation}
where the internal temperatures $T_i$ are obtained from microscopic entropy functions $s_i(\eps_i)$ (which can be different for different particles, as emphasized by the notation) as 
\[
T_i(\eps) = \frac{1}{k_{\rm B}s_i'(\eps)}.
\] 
See Section~\ref{sec:microEOS} below for further precisions on the micro-equation of state (EOS) $s_i(\eps_i)$. The choice~\eqref{eq:FDR1} ensures that, for any given energy level $E_0 \geq -\min V$, the following measure is invariant by the dynamics:
\begin{equation}
  \label{eq:nu_N}
  \nu_N(dq\,dp\,d\eps) = Z_{\nu,N}^{-1} \prod_{i=1}^N \rme^{s_i(\eps_i)} \delta_{\{ \mathcal{E}(q,p,\eps) - E_0 \}}(dq\,dp\,d\eps). 
\end{equation}
Here again, this preservation is ensured by the fact that each elementary dynamics preserves~$\nu_N$; see again Section~\ref{sec:splitting}. The invariant measure $\nu_N$ is equivalent, in the thermodynamic limit, to the canonical measure
\begin{equation}
  \label{eq:canonical_N}
  \mu_N(dq\,dp\,d\eps) = Z_\beta^{-1} \, \rme^{-\beta H(q,p)} \, \prod_{i=1}^N \rme^{s_i(\eps_i)-\beta \eps_i} \, d\eps_i \, dq \, dp,
\end{equation}
where $\beta = 1/(\kB T_\beta)$ is such that the average energy under the canonical measure is equal to the prescribed energy level: $\langle \cE \rangle_{\mu_N} = E_0$. 

\subsection{Micro-equation of state}
\label{sec:microEOS}

One key ingredient in DPDE is the micro-EOS which relates the entropy and the internal energy. With some abuse of notation, any of the internal energies $\eps_1,\dots,\eps_N$ is simply denoted by $\eps$ in this section. In general, the internal temperature 
\begin{equation}
  \label{eq:def_temp_microEOS}
  T(\eps) = \frac{1}{\kB s'(\eps)}
\end{equation}
associated with an internal energy $\eps$ is implicitly defined from the internal energy~$\eps$ via the relation
\begin{equation}
  \label{eq:eps_integral_Cv}
  \eps = \int_0^{T(\eps)} C_v(\theta) \, d\theta,
\end{equation}
where $C_v(\theta)$ is the (temperature-dependent) heat capacity. Moreover, the marginal of the canonical measure~\eqref{eq:canonical_N} in the variable~$\eps$ reads
\begin{equation}
  \label{eq:marginal_canonical_eps}
  \mu_{\rm int}(d\eps) = Z_\eps^{-1} \exp(s(\eps)-\beta\eps) \, d\eps.
\end{equation}

\subsubsection{Classical micro-EOS}

The classical micro-EOS corresponds to a constant heat capacity, in which case $T(\eps) = \eps/\Cv$ and
\begin{equation}
\label{eq:s_classical}
s(\eps) = \frac{\Cv}{\kB} \ln\left(\frac{\eps}{\epsref}\right),
\end{equation}
where $\epsref > 0$ is some reference energy. More realistic models, fitted on ab-initio simulations, require a genuinely temperature-dependent heat capacity. Stability issues for the numerical integrators may be magnified in these cases. The next sections introduce empirical models taking into account some temperature dependence, which are relevant to test the robustness of the numerical scheme for general micro-EOS. 

\subsubsection{Einstein model}

\begin{figure} 
\includegraphics[width=0.5\textwidth]{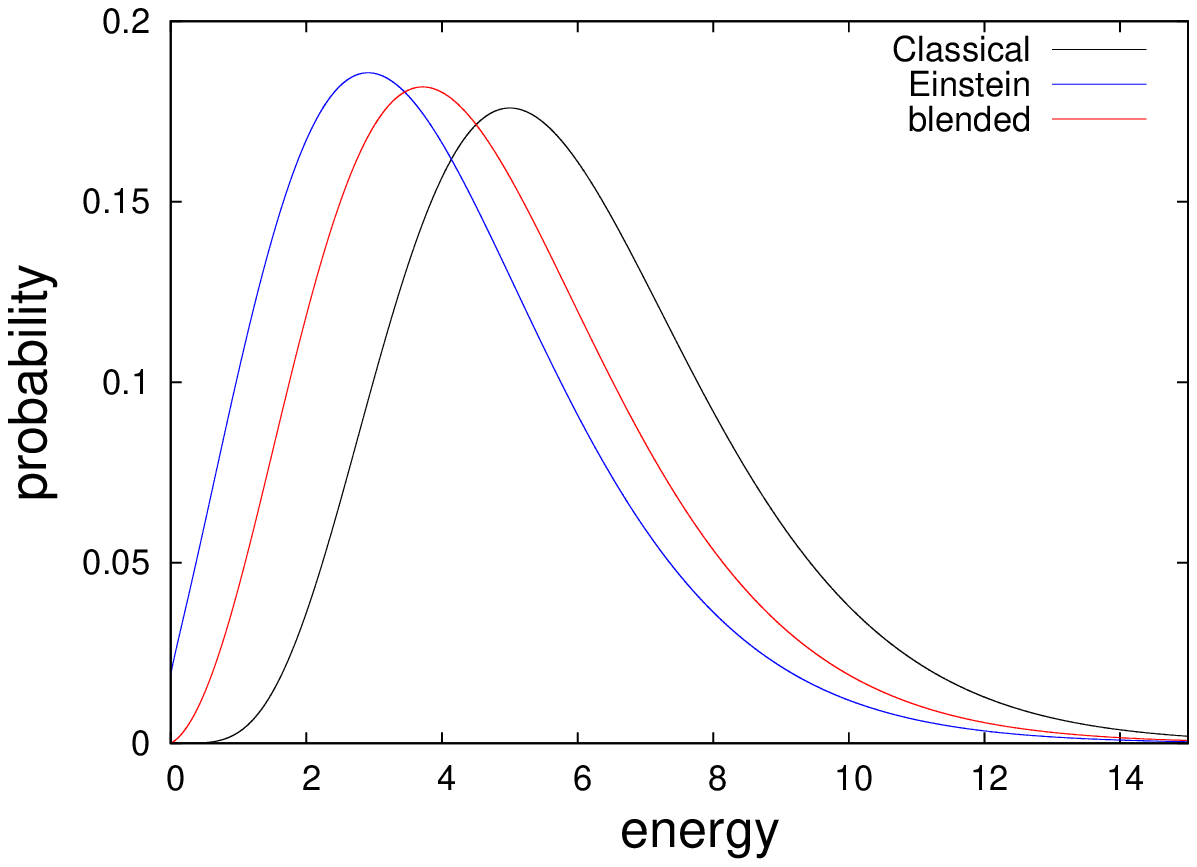}
\includegraphics[width=0.5\textwidth]{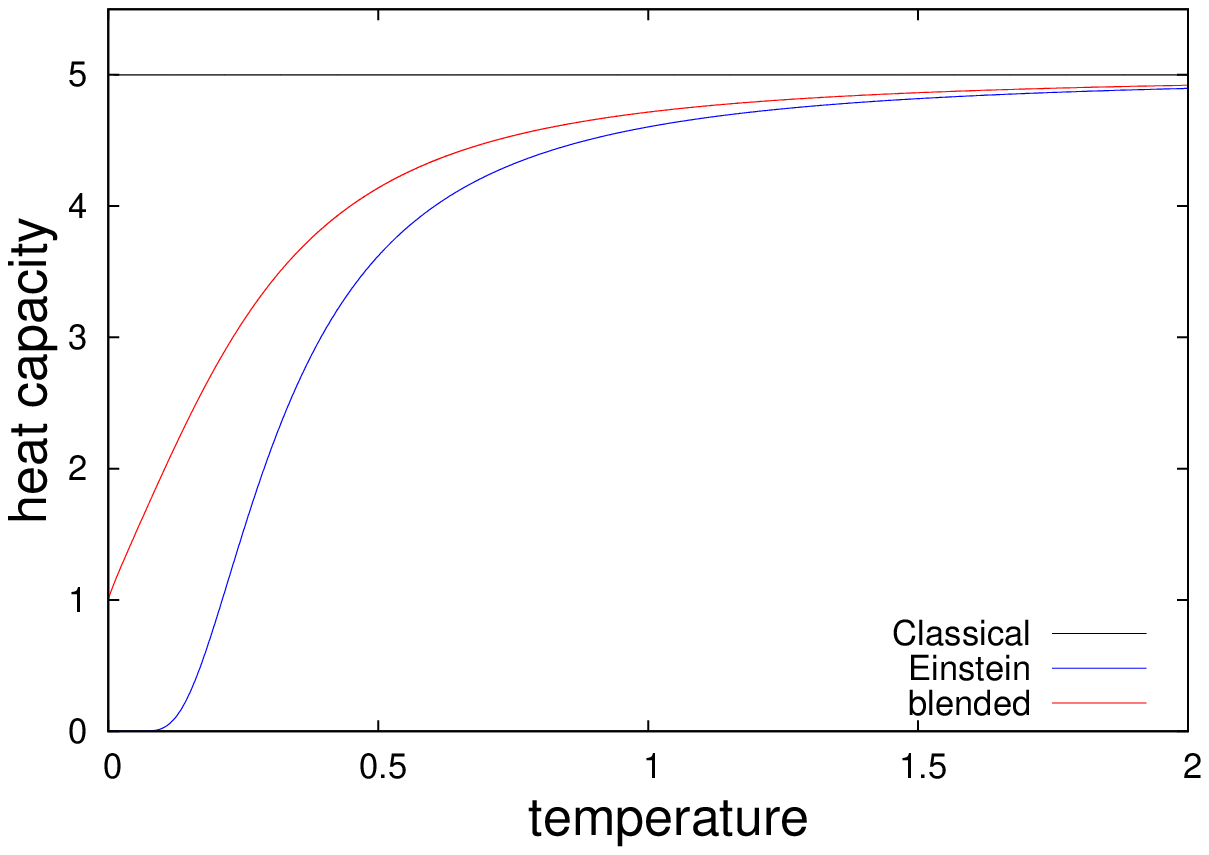}
\caption{Left: Probability distribution~\eqref{eq:marginal_canonical_eps} at $\beta = 1$, for various choices of entropy functions (in reduced units where $\epsref = 1$ and $\kB = 1$). Right: Associated heat capacities. 'Classical' refers to~\eqref{eq:s_classical} with $\Cv = 5$, 'Einstein' to~\eqref{eq:s_Einstein} with $\Cv = 5$ and $\To = 1$, 'blended' to~\eqref{eq:s_blended} with $\Cv = 5$, $\Cvo = 1$ and $\To = 1$.}
\label{fig:Einstein}
\end{figure}

A first model for temperature-dependent heat capacities is obtained from the Einstein model of harmonic oscillators, already considered in~\cite{KSM16}. A full derivation of the equations presented in this section is given in Appendix~A. The Einstein model corresponds to the following micro-EOS:
\begin{equation}
\label{eq:s_Einstein}
s(\eps) = \frac{1}{\kB\To} \left[ (\eps + \Cv\To) \ln \left(\frac{\eps + \Cv\To}{\epsref}\right) - \eps \ln\left(\frac{\eps}{\epsref}\right) \right],
\end{equation}
where $\To$ is some reference temperature, and $\Cv$ the limiting heat capacity for large temperatures. The classical micro-EOS~\eqref{eq:s_classical} is recovered in the limit $\To \to 0$ (up to an unimportant additive constant). The internal temperature associated with~\eqref{eq:s_Einstein} reads (see~\eqref{eq:temp_fct_eps})
\[
T(\eps) = \frac{1}{\kB s'(\eps)} = -\frac{\To}{\dps \ln\left(1 - \frac{\Cv \To}{\eps + \Cv\To}\right)},
\]
while the associated heat capacity is (see~\eqref{eq:Cv_theta_Einstein})
\[
C_v(\theta) = \Cv \, \left(\frac{\To}{\theta}\right)^2 \frac{\rme^{-\To/\theta}}{\left(1 - \rme^{-\To/\theta}\right)^2}.
\]
See Figure~\ref{fig:Einstein} for plots of the associated distribution of internal energies, and of the heat capacity as a function of the temperature. Note that, in practice, only $s(\eps)$ and $s'(\eps)$ are needed to integrate the dynamics. The heat capacity is useful only for physical interpretation.

An important point is that the definition of the thermodynamic temperature has to be changed since $s(\eps)$ does not tend to $-\infty$ as $\eps \to 0$. One estimator of the internal temperature is for instance
\begin{equation}
  \label{eq:estimator_temperature_eps}
  T_\beta = \frac{\langle \eps \rangle}{\dps \kB\left(1 + \langle s'(\eps) \eps \rangle\right)},
\end{equation}
where $\langle \cdot\rangle$ refers to averages with respect to the marginal measure~\eqref{eq:marginal_canonical_eps}. In fact, there is whole family of temperature estimators, see~\eqref{eq:estimator_temperature_Einstein} in Appendix~A for further precisions.

\subsubsection{Blended Einstein model}
\label{sec:blended}

The internal entropy in the Einstein model is such that the marginal measures $Z_\eps^{-1} \rme^{s(\eps)-\beta \eps} \, d\varepsilon$ have non-zero finite values at $\eps = 0$. This may lead to thermodynamic inconsistencies in the model. It seems more appropriate to consider a model micro-EOS which ensures that the marginal measure vanishes at $\eps = 0$, so that negative energies cannot appear for the continuous dynamics. The model EOS is obtained by adding an Einstein contribution (of maximal value $\Cv-\Cvo$) to a baseline constant heat capacity $\Cvo$. More precisely,
\begin{equation}
\label{eq:s_blended}
s(\eps) = \frac{\Cvo}{\kB} \ln\left(\frac{\eps}{\epsref}\right) + \frac{1}{\kB\To} \left[ (\eps + (\Cv-\Cvo)\To) \ln \left(\frac{\eps + (\Cv-\Cvo)\To}{\epsref}\right) - \eps \ln\left(\frac{\eps}{\epsref}\right) \right].
\end{equation}
Note that that the corresponding heat capacity $C_v(\theta)$ is such that $C_v(0) = \Cvo$ and $C_v(\theta) \to \Cv$ as $\theta \to +\infty$. For this model, the standard internal temperature estimator 
\[
T_\beta = \left\langle \frac{1}{T(\eps)} \right\rangle^{-1}
\]
can therefore be used (see~\eqref{eq:std_int_temp} in Appendix~A). Figure~\ref{fig:Einstein} shows a typical distribution of internal energies and the heat capacity associated with this model. Since there is no closed form expression for the latter quantity, the picture has been obtained by plotting $1/T'(\eps)$ as a function of $T(\eps)$, relying on~\eqref{eq:Cv_parametric_curve}.

\subsection{Integration by splitting}
\label{sec:splitting}

When the timestep $\dt > 0$ is fixed, a numerical integration of stochastic dynamics such as DPDE consists in finding an iteration rule to pass from $(q^n,p^n,\eps^n)$, an approximation of the solution $(q_{n\dt},p_{n\dt},\eps_{n\dt})$ of~\eqref{eq:DPDE} at time $n\dt$, to an approximation at the next timestep, namely $(q^{n+1},p^{n+1},\eps^{n+1})$. The strategy developped here consists in integrating successively the various subparts of the dynamics. 

\subsubsection{Decomposition into elementary dynamics}

DPDE can be decomposed into several elementary dynamics:
\begin{enumerate}[(i)] 
\item The first one is the Hamiltonian part
\[
\left\{ \begin{aligned}
  dq_{i} & = \frac{p_{i,t}}{m_i} \, dt, \\
  dp_{i} & = -\nabla_{q_i} V(q) \, dt,
\end{aligned} \right.
\]
which preserves the mechanical energy $H(q,p)$ (and hence the total energy~$\cE$), as well as the measure~\eqref{eq:nu_N}. 
\item The second family of elementary dynamics are the pairwise fluctuation/dissipation dynamics
\begin{equation}
\label{eq:DPDE_elementary}
\left \{
\begin{aligned}
dp_i & = -\gamma(\eps_i,\eps_j) \chi^2(r_{ij}) \Big(e_{ij} \cdot v_{ij} \Big) e_{ij} \, dt + \sigma \chi(r_{ij}) e_{ij} \, dW_{ij}, \\
dp_j & = -dp_i, \\
d\eps_j & = \frac{\chi^2(r_{ij})}{2} \left[ \gamma(\eps_i,\eps_j)\Big(e_{ij} \cdot v_{ij} \Big)^2 - \frac{\sigma^2}{2} \left(\frac{1}{m_i}+\frac{1}{m_j}\right)\right]dt - \frac{\sigma}{2} \chi(r_{ij}) \Big(v_{ij} \cdot e_{ij}\Big) dW_{ij}, \\
d\eps_j & = d\eps_i.
\end{aligned}
\right.
\end{equation}
The evolution of $\eps_i,\eps_j$ is in fact fully determined by the requirements that $d\eps_i = d\eps_j$ and the elementary energy
\[
\cE_{ij}(p_i,p_j,\eps_i,\eps_j) = \frac{p_i^2}{2m_i} + \frac{p_j^2}{2m_j} + \eps_i + \eps_j  
\]
be constant (using some It\^o calculus). It can also be shown that the elementary dynamics~\eqref{eq:DPDE_elementary} preserves any measure of the form $\rme^{s_i(\eps_i)+s_j(\eps_j)} f(\cE_{ij}) \,dp_i \, dp_j\, d\eps_i \, d\eps_j$, so that it preserves in particular the measure~\eqref{eq:nu_N} and the total energy~$\cE$. 
\item The third and last family of elementary dynamics are the pairwise elementary conduction dynamics 
\begin{equation}
\label{eq:DPDE_elementary_cond}
\left\{ \begin{aligned}
d\eps_i & = \kappa \chi^2(r_{ij}) \left(\frac{1}{T_i(\eps_i)}-\frac{1}{T_j(\eps_j)} \right) dt + \sqrt{2\kappa}\chi(r_{ij}) \, d\widetilde{W}_{ij}, \\
d\eps_j & = -d\eps_i.
\end{aligned} \right.
\end{equation}
By construction, these subdynamics leave the energy $\eps_i+\eps_j$ invariant, hence the total energy $\cE$ as well. In addition, it can also be shown that any measure of the form $\rme^{s_i(\eps_i)+s_j(\eps_j)} f(\eps_i+\eps_j) \, d\eps_i \, d\eps_j$ is invariant, so that~\eqref{eq:DPDE_elementary_cond} preserves in particular the measure~\eqref{eq:nu_N}. 
\end{enumerate}

\subsubsection{Splitting schemes}

In view of the above physical decomposition of the full DPDE, a numerical integrator can be obtained by composing integrators for all elementary dynamics under consideration -- a strategy known as splitting methods. For the Hamiltonian part, the standard choice is to use the Verlet scheme~\cite{Verlet}, which corresponds to the integrator $(q^{n+1},p^{n+1}) = \Phi_\dt^{\rm Verlet}(q^n,p^n)$ with
\[
\begin{aligned}
& \Phi_\dt^{\rm Verlet}(q,p) \\
& = \left(q+\dt M^{-1}p-\frac{\dt^2}{2}\nabla V(q), p-\frac{\dt}{2}\left[\nabla V(q)+\nabla V\left(q+\dt M^{-1}p-\frac{\dt^2}{2}\nabla V(q)\right)\right]\right).
\end{aligned}
\]
Integrators for the elementary dynamics~\eqref{eq:DPDE_elementary} and~\eqref{eq:DPDE_elementary_cond} are respectively denoted by 
\[
\Phi_\dt^{{\rm FD},ij}\left(p_i,p_j,\eps_i,\eps_j,G_{ij},U_{ij}\right), 
\qquad
\Phi_\dt^{{\rm TC},ij}\left(\eps_i,\eps_j,\widetilde{G}_{ij},\widetilde{U}_{ij}\right),
\]
where 'FD' stands for fluctuation/dissipation and 'TC' for thermal conduction. Note that these integrators depend on certain random numbers: independent Gaussian random variables $G_{ij}$ and $\widetilde{G}_{ij}$ to discretize the Brownian motions $W_{ij}$ and $\widetilde{W}_{ij}$, as well as uniform random variables $U_{ij}$ and $\widetilde{U}_{ij}$ which are used to implement a Metropolis correction. If a standard discretization of~\eqref{eq:DPDE_elementary} and~\eqref{eq:DPDE_elementary_cond} is considered (using, say, stochastic Runge--Kutta methods), then no uniform random variable is needed; on the other hand, several Gaussian variables may be required to integrate the dynamics over one timestep. See for instance~\cite{MT04} for an introduction to numerical schemes for SDEs.

One possible scheme is the following. Denoting by $r_{\rm cut}$ the range of the cut-off function~$\chi$, the set of ``active'' pairs for a given set of positions~$q$ (\textit{i.e.} the set of pairs for which the elementary dynamics~\eqref{eq:DPDE_elementary} and~\eqref{eq:DPDE_elementary_cond} are not trivial) is
\[
\mathcal{A}(q) = \left\{ (i,j) \in \{1,\dots,N\}^2 \, \big| \, i < j, \ |q_i-q_j| \leq r_{\rm cut} \right\}.
\]
A new configuration $(q^{n+1},p^{n+1},\eps^{n+1})$ is then obtained from $(q^n,p^n,\eps^n)$ for instance by the composition
\begin{equation}
\label{eq:splitting_scheme}
(q^{n+1},p^{n+1},\eps^{n+1}) = \left(\mathop{\bigcirc}_{(i,j)\in\mathcal{A}(q^{n+1})}\Phi_\dt^{{\rm TC},ij}\right)\left(\mathop{\bigcirc}_{(i,j)\in\mathcal{A}(q^{n+1})}\Phi_\dt^{{\rm FD},ij}\right) \circ \Phi_\dt^{\rm Verlet}(q^n,p^n,\eps^n),
\end{equation}
which corresponds to first integrating the Hamiltonian dynamics with the Verlet scheme, then looping over the active pairs to integrate the fluctuation/dissipation, and finally looping again over the active pairs to integrate the thermal conduction. Several comments are in order on this formula. Note first the abuse of notation which consists in not making explicit the actual variables of the various integrators (sometimes additional variables are considered, such as $\eps^n$ for the Verlet scheme; while the random variables are omitted). Second, note that the only scheme which modifies positions is the Verlet scheme, which is why the active pairs are determined based on $q^{n+1}$, the positions coming out of $\Phi_\dt^{\rm Verlet}$. Last, let us emphasize that the order of integration is somewhat arbitrary: it is equally possibly to finish by the Verlet part, and/or to immediately integrate both fluctuation/dissipation and thermal conduction for a given pair in order to avoir looping twice over pairs; maybe more importantly, it is difficult, if not impossible on modern computing architectures, to assign an order to the way pairs are looped over (based on, say, lexicographical order): when DPDE is parallelized as in~\cite{LBMLM14}, the order is determined by the spatial decomposition under consideration.

Error estimates on average properties can be deduced from the integration errors on each subdynamics, using an analysis similar to the one used for Langevin dynamics in~\cite{LMS16}. This analysis is made precise in Section~\ref{sec:errors}, after a description of the integrators $\Phi_\dt^{{\rm FD},ij}$ and $\Phi_\dt^{{\rm TC},ij}$. 

\subsubsection{Multiple timestep strategies}
\label{sec:MTS}

As will be made clear in the numerical examples presented in Section~\ref{sec:numerics}, the stability of splitting schemes such as~\eqref{eq:splitting_scheme} is limited in practice by the Hamiltonian part of the dynamics, especially when singular interaction potentials (\textit{e.g.} Lennard--Jones or Coulomb) are considered. One option in this case is to resort to multiple timestepping strategies, where the Hamiltonian part is integrated with a smaller timestep. The reference timestep for the integration of the Hamiltonian part is denoted by $\Delta t_{\rm Ham}$. Introducing an integer $k_{\rm MTS} \geq 1$, the timestep used to integrate the elementary pairwise stochastic interactions is $\dt = k_{\rm MTS} \Delta t_{\rm Ham}$. This amounts to replacing the integrator in~\eqref{eq:splitting_scheme} with
\begin{equation}
\label{eq:splitting_scheme_MTS}
\left(\mathop{\bigcirc}_{(i,j)\in\mathcal{A}(q^{n+1})}\Phi_\dt^{{\rm TC},ij}\right)\left(\mathop{\bigcirc}_{(i,j)\in\mathcal{A}(q^{n+1})}\Phi_\dt^{{\rm FD},ij}\right) \circ \left(\Phi_{\dt_{\rm Ham}}^{\rm Verlet}\right)^{k_{\rm MTS}}.
\end{equation}
Note that this scheme still provides a consistent discretization of the original dynamics when $k_{\rm MTS}$ is fixed and $\dt_{\rm Ham} \to 0$. 

\section{Integrating elementary pairwise stochastic interactions}
\label{sec:metropolis}

This section presents stable schemes to integrate the elementary pairwise fluctuation/dissipation and thermal conduction dynamics. The key idea, made precise in Section~\ref{sec:rewriting} for the fluctuation/dissipation and in Section~\ref{sec:num_TC} for thermal conduction, is to rewrite the elementary dynamics as effective Brownian dynamics of a single variable. The invariant measure of these dynamics is analytically known in terms of the state of the system at step~$n$, which allows to correct numerical discretizations by a Metropolis--Hastings procedure (described in Sections~\ref{sec:Metropolis_elementary_FD} and~\ref{sec:num_TC}). The Metropolis correction both allows to stabilize numerical schemes by automatically rejecting negative energies, and also prevents any bias on the thermodynamic properties. The error on average properties for the resulting numerical scheme therefore solely arises from the Verlet discretization, as made precise in Section~\ref{sec:errors}.

\subsection{Rewriting fluctuation/dissipation dynamics as effective single-variable dynamics}
\label{sec:rewriting}

For notational simplicity, consider the elementary fluctuation/dissipation dynamics associated with particles~1 and~2 (rather than general indices~$i$ and~$j$):
\begin{equation}
\label{eq:DPDE_elementary_12}
\left \{
\begin{aligned}
dp_1 & = -\gamma(\eps_1,\eps_2) \chi^2(r_{12}) \Big(e_{12} \cdot v_{12} \Big) e_{12} \, dt + \sigma \chi(r_{12}) e_{12} \, dW_t, \\
dp_2 & = -dp_1, \\
d\eps_1 & = \frac{\chi^2(r_{12})}{2} \left[ \gamma(\eps_1,\eps_2)\Big(e_{12} \cdot v_{12} \Big)^2 - \frac{\sigma^2}{2} \left(\frac{1}{m_1}+\frac{1}{m_2}\right)\right]dt - \frac{\sigma}{2} \chi(r_{12}) \Big(v_{12} \cdot e_{12}\Big) dW_t, \\
d\eps_2 & = d\eps_1,
\end{aligned}
\right.
\end{equation}
where $W_t$ is a standard one-dimensional Brownian motion. Recall that the evolution of $\eps_1,\eps_2$ is in fact fully determined by the requirement that the energy
\begin{equation}
\label{eq:elementary_energy}
\mathscr{E}(p_1,p_2,\eps_1,\eps_2) = \frac{p_1^2}{2m_1} + \frac{p_2^2}{2m_2} + \eps_1 + \eps_2  
\end{equation}
be constant. It is therefore sufficient to integrate the dynamics on $p_1$, from which the evolution of all other variables (namely $p_2,\eps_1,\eps_2$) is deduced. Recall also that the derivation presented here and in Section~\ref{sec:Metropolis_elementary_FD} for elementary fluctuation/dissipation dynamics projected along lines of center are generalized in Appendix~B.

In order obtain a simplified elementary dynamics, note first that the components of $p_1,p_2$ orthogonal to $e_{12}$ do not evolve in time, and that $p_1+p_2$ is conserved. It is therefore sufficient to determine the evolution of the relative velocity along the lines of centers, namely $v_{12}^\pr  = v_{12} \cdot e_{12} \in \mathbb{R}$ where 
\[
v_{12} = \frac{p_1}{m_1} - \frac{p_2}{m_2}.
\]
The projection of~\eqref{eq:DPDE_elementary_12} onto the direction~$e_{12}$ leads to the following equation for $v_{12}^\pr$: 
\begin{equation}
  \label{eq:eq_v12_pr}
  dv_{12}^\pr = - \gamma(\eps_1,\eps_2) \chi^2(r_{12}) \left(\frac{1}{m_1} + \frac{1}{m_2}\right) v_{12}^\pr \, dt + \sigma\chi(r_{12}) \left(\frac{1}{m_1} + \frac{1}{m_2}\right)dW_t.
\end{equation}

The second observation is that the energy~\eqref{eq:elementary_energy} is preserved, which, together with the last two lines of~\eqref{eq:DPDE_elementary_12}, implies that
\begin{equation}
\label{eq:internal_energies_in_terms_of_p}
\begin{aligned}
\eps_1 = \eps_{1,0} + \frac12\left( \frac{p_1^2 - p_{1,0}^2}{2m_1} + \frac{p_2^2 - p_{2,0}^2}{2m_2}\right), \qquad
\eps_2 = \eps_{2,0} + \frac12\left( \frac{p_1^2 - p_{1,0}^2}{2m_1} + \frac{p_2^2 - p_{2,0}^2}{2m_2}\right),
\end{aligned}
\end{equation}
where the quantities with subscripts~0 indicate initial conditions while quantities without subscripts implicitly indicate values at time~$t > 0$. Now, the momenta $p_1,p_2$ can be expressed in terms of their initial values and the current value of the relative velocity $v_{12}^\pr$. A simple computation shows that
\begin{equation}
\label{eq:momenta_in_terms_of_v}
\begin{aligned}
p_{1} & = \left(\frac{1}{m_1} + \frac{1}{m_2}\right)^{-1}\left(\frac{p_{1,0}+p_{2,0}}{m_2} + v_{12,0}^\perp + v_{12}^\pr e_{12} \right) = p_{1,0} + \left(\frac{1}{m_1} + \frac{1}{m_2}\right)^{-1}\left(v_{12}^\pr - v_{12,0}^\pr\right)e_{12}, \\
p_{2} & = \left(\frac{1}{m_1} + \frac{1}{m_2}\right)^{-1}\left(\frac{p_{1,0}+p_{2,0}}{m_1} - v_{12,0}^\perp - v_{12}^\pr e_{12}\right) = p_{2,0} - \left(\frac{1}{m_1} + \frac{1}{m_2}\right)^{-1}\left(v_{12}^\pr - v_{12,0}^\pr\right)e_{12},
\end{aligned}
\end{equation}
where it was made use of the fact that $p_{1,0}+p_{2,0}$ and $v_{12,0}^\perp = v_{12,0} - (v_{12,0}\cdot e_{12})e_{12}$ are invariants of the elementary dynamics~\eqref{eq:DPDE_elementary_12}. For more compact notation, introduce the reduced mass
\[
\mu_{12} = \left(\frac{1}{m_1} + \frac{1}{m_2}\right)^{-1}.
\]
Plugging~\eqref{eq:momenta_in_terms_of_v} into~\eqref{eq:internal_energies_in_terms_of_p} allows to write the internal energies as a function of $v_{12}^\pr$. Since
\[
\frac{p_1^2 - p_{1,0}^2}{2m_1} = \mu_{12} \left(v_{12}^\pr - v_{12,0}^\pr\right) \frac{p_{1,0}}{m_1}\cdot e_{12} + \frac{\mu_{12}^2}{2m_1}\left(v_{12}^\pr - v_{12,0}^\pr\right)^2,
\]
it follows that
\begin{equation}
\label{eq:kin_energy_difference}
\begin{aligned}
\frac{p_1^2 - p_{1,0}^2}{2m_1} + \frac{p_2^2 - p_{2,0}^2}{2m_2} & = \mu_{12} \left(v_{12}^\pr - v_{12,0}^\pr\right) v_{12,0}^\pr + \frac{\mu_{12}}{2}\left(v_{12}^\pr - v_{12,0}^\pr\right)^2 \\
& = \frac{\mu_{12}}{2} \left[ \left(v_{12}^\pr\right)^2 - \left(v_{12,0}^\pr\right)^2 \right]. 
\end{aligned}
\end{equation}
Therefore, 
\begin{equation}
\label{eq:eps_in_terms_of_v}
\eps_1 = \eps_{1,0} - \frac{\mu_{12}}{4} \left[ \left(v_{12}^\pr\right)^2 - \left(v_{12,0}^\pr\right)^2 \right], 
\qquad
\eps_2 = \eps_{2,0} - \frac{\mu_{12}}{4} \left[ \left(v_{12}^\pr\right)^2 - \left(v_{12,0}^\pr\right)^2 \right]. 
\end{equation}

By plugging the expressions of the internal energies into~\eqref{eq:eq_v12_pr}, the effective one-dimensional dynamics on $v_{12}^\pr$ finally reads
\begin{equation}
\label{eq:eff_dyn}
dv_{12}^\pr = - \Gamma\left(v_{12}^\pr\right) \chi^2(r_{12}) \left(\frac{1}{m_1} + \frac{1}{m_2}\right) v_{12}^\pr \, dt + \sigma\chi(r_{12}) \left(\frac{1}{m_1} + \frac{1}{m_2}\right)dW_t,
\end{equation}
with
\begin{equation}
  \label{eq:Gamma}
  \Gamma(v) = \frac{\sigma^2}{4k_{\rm B}}\left\{ s_1'\left(\eps_{1,0} - \frac{\mu_{12}}{4} \left[ v^2 - \left(v_{12,0}^\pr\right)^2 \right]\right) + s_2'\left(\eps_{2,0} - \frac{\mu_{12}}{4} \left[ v^2 - \left(v_{12,0}^\pr\right)^2 \right]\right) \right\}.
\end{equation}
The effective dynamics~\eqref{eq:eff_dyn} is the reference dynamics upon which the numerical integrator $\Phi_\dt^{{\rm FD},12}$ is constructed. Note that it is parametrized by the initial conditions $\eps_{1,0},\eps_{2,0}$ and $v_{12,0}^\pr$. In the numerical scheme presented in Section~\ref{sec:Metropolis_elementary_FD}, these initial conditions are the values at iteration~$n$, while the current values of the effective dynamics~\eqref{eq:eff_dyn} at time~$\dt$ provide the values at the next iteration~$n+1$. 

The effective dynamics~\eqref{eq:eff_dyn} can be rewritten as an overdamped Langevin dynamics as follows:
\[
dv_{12}^\pr = - \frac12 B(r_{12})^2 \mathcal{U}\left(v_{12}^\pr\right)\,dt + B(r_{12}) \, dW_t,
\]
with
\[
B(r_{12}) = \frac{\sigma \chi(r_{12})}{\mu_{12}},
\]
and 
\[
\mathcal{U}(v) = \frac{\mu_{12}}{2}\left\{ s_1'\left(\eps_{1,0} - \frac{\mu_{12}}{4} \left[ v^2 - \left(v_{12,0}^\pr\right)^2 \right]\right) + s_2'\left(\eps_{2,0} - \frac{\mu_{12}}{4} \left[ v^2 - \left(v_{12,0}^\pr\right)^2 \right]\right) \right\} v.
\]
Note that $\mathcal{U} = U'$ with
\[
U(v) = -s_1\left(\eps_{1,0} - \frac{\mu_{12}}{4} \left[ v^2 - \left(v_{12,0}^\pr\right)^2 \right]\right) - s_2\left(\eps_{2,0} - \frac{\mu_{12}}{4} \left[ v^2 - \left(v_{12,0}^\pr\right)^2 \right]\right).
\]
When $B(r_{12})>0$, the unique invariant measure of~\eqref{eq:eff_dyn} therefore reads
\begin{equation}
\label{eq:inv_meas_reduced_dyn_v}
\begin{aligned}
\nu(dv) & = Z_\nu^{-1} \rme^{-U(v)} \, dv \\
& = Z_\nu^{-1} \exp\left[s_1\left(\eps_{1,0} - \frac{\mu_{12}}{4} \left[ v^2 - \left(v_{12,0}^\pr\right)^2 \right]\right) + s_2\left(\eps_{2,0} - \frac{\mu_{12}}{4} \left[ v^2 - \left(v_{12,0}^\pr\right)^2 \right]\right)\right] \, dv.
\end{aligned}
\end{equation}

\subsection{Metropolization of elementary fluctuation/dissipation dynamics}
\label{sec:Metropolis_elementary_FD}

To simplify the notation, $v_{12}^\pr$ is replaced by $v$ in this section. The proposed numerical scheme consists in (i) proposing a new move by analytically integrating the effective dynamics~\eqref{eq:eff_dyn} over a time~$\dt$, with initial conditions $p_1^n,p_2^n,\eps_1^n,\eps_2^n$ and with the friction fixed to $\gamma^n := \gamma(\eps_1^n,\eps_2^n) = \Gamma(v_{12}^n)$; then (ii) accepting or rejecting this proposal according a Metropolis criterion. This corresponds to the so-called SmartMC algorithm~\cite{RDF78}. More precisely, the proposed new velocity is
\begin{equation}
\label{eq:exp_integrator}
\widetilde{v}^{n+1} = \alpha^n v^n +  \eta^n \, G^n,
\end{equation}
where $G^n$ is a sequence of independent and identically distributed standard one-dimensional Gaussian random variables, and 
\[
\alpha^n = \exp\left(-\gamma^n \frac{\chi(r_{12})^2}{\mu_{12}}\dt \right), 
\qquad
\eta^n = \sigma \sqrt{\frac{1-(\alpha^n)^2}{2 \gamma^n \mu_{12}}}.
\]
The new momenta and internal energies are then obtained from~\eqref{eq:momenta_in_terms_of_v} and~\eqref{eq:eps_in_terms_of_v}. This scheme in fact coincides with the SSA discretization considered in~\cite{HMS16}. The difference with the standard SSA scheme is however that~\eqref{eq:exp_integrator} only provides a proposal for the new state, to be accepted or rejected.

The Metropolis ratio to accept a proposed transition from $v$ to $v'$ is $\min\left(1,A_\dt(v,v')\right)$, with
\[
A_\dt(v,v') = \frac{\nu(v')T_\dt(v',v)}{\nu(v)T_\dt(v,v')},
\]
where $T_\dt(v,v')$ is the transition kernel associated with the numerical scheme and $\nu$ is defined in~\eqref{eq:inv_meas_reduced_dyn_v} (upon replacing $\eps_{1,0},\eps_{2,0}$ with $\eps_1^n,\eps_2^n$). For~\eqref{eq:exp_integrator}, it holds
\begin{equation}
\label{eq:transition_kernel_dt}
T_\dt(v^n,\widetilde{v}^{n+1}) = \frac{1}{\eta^n \sqrt{2\pi}}\exp\left(-\frac{|\widetilde{v}^{n+1}-\alpha^n v^n|^2}{2(\eta^n)^2}\right) = \frac{1}{\eta^n \sqrt{2\pi}}\exp\left(-\frac{|G^n|^2}{2}\right).
\end{equation}
In addition, using the configuration at iteration~$n$ as the reference in~\eqref{eq:inv_meas_reduced_dyn_v}, and denoting by $\eps_1^n,\eps_2^n$ the internal energies at this time,
\[
\log\left( \frac{\nu(\widetilde{v}^{n+1})}{\nu(v^n)} \right) = s_1\!\left(\widetilde{\eps}_1^{n+1}\right) + s_2\!\left(\widetilde{\eps}_2^{n+1}\right)- s_1(\eps_1^n) - s_2(\eps_2^n),
\]
where
\[
\widetilde{\eps}_1^{n+1} = \eps_1^n - \frac{\mu_{12}}{4}\left[\left(\widetilde{v}^{n+1}\right)^2 - \left(v_{12}^n\right)\right],
\qquad
\widetilde{\eps}_2^{n+1} = \eps_2^n - \frac{\mu_{12}}{4}\left[\left(\widetilde{v}^{n+1}\right)^2 - \left(v_{12}^n\right)\right].
\]
If one of the proposed new energies $\widetilde{\eps}_1^{n+1},\widetilde{\eps}_2^{n+1}$ is negative, $\nu(\widetilde{v}^{n+1})$ is set to~0 to avoid singularities (in a subsequent step of the algorithm, these configurations are anyway automatically discarded). When the new proposed energies are positive, the probability of the reverse move starting from $\widetilde{v}^{n+1},\widetilde{\eps}_1^{n+1},\widetilde{\eps}_2^{n+1}$ is needed. Denoting by $\widetilde{\gamma}^{n+1} = \gamma(\widetilde{\eps}_1^{n+1},\widetilde{\eps}_2^{n+1})$ the friction associated with $\widetilde{\eps}_1^{n+1},\widetilde{\eps}_2^{n+1}$, and introducing
\[
\widetilde{\alpha}^{n+1} = \exp\left(- \widetilde{\gamma}^{n+1}\frac{\chi(r_{12})^2}{\mu_{12}}\dt \right), 
\qquad
\widetilde{\eta}^{n+1} = \sigma \sqrt{\frac{1-(\widetilde{\alpha}^{n+1})^2}{2 \widetilde{\gamma}^{n+1} \mu_{12}}},
\]
it holds
\[
T_\dt(\widetilde{v}^{n+1},v^n) = \frac{1}{\widetilde{\eta}^{n+1} \sqrt{2\pi}}\exp\left(-\frac{|v^n-\widetilde{\alpha}^{n+1}\widetilde{v}^{n+1}|^2}{2(\widetilde{\eta}^{n+1})^2}\right).
\]
The complete expression of the acceptance therefore relies on the following quantity:
\begin{equation}
\label{eq:expression_A_dt}
\begin{aligned}
a_\dt(v^n,\widetilde{v}^{n+1}) := \log A_\dt(v^n,\widetilde{v}^{n+1}) & = s_1\!\left(\widetilde{\eps}_1^{n+1}\right) + s_2\!\left(\widetilde{\eps}_2^{n+1}\right)- s_1(\eps_1^n) - s_2(\eps_2^n) \\
& \ \ + \frac{(G^n)^2}{2} + \log \eta^n - \frac{|v^n-\widetilde{\alpha}^{n+1}\widetilde{v}^{n+1}|^2}{2(\widetilde{\eta}^{n+1})^2} - \log \widetilde{\eta}^{n+1}.
\end{aligned}
\end{equation}

The precise algorithm to integrate elementary dynamics such as~\eqref{eq:DPDE_elementary} is the following. Starting from a current configuration $(p_1^n,p_2^n,\eps_1^n,\eps_2^n)$:
\begin{enumerate}[(i)]
\item compute $\dps v^n = \left(\frac{p^n_1}{m_1}-\frac{p^n_2}{m_2}\right)\cdot e_{12}$;
\item propose a new value $\widetilde{v}^{n+1}$ according to~\eqref{eq:exp_integrator};
\item check whether the following energy bound is satisfied: 
\[
\frac{\mu_{12}}{4} \left[ \left(\widetilde{v}^{n+1}\right)^2 - \left(v^n\right)^2 \right] \leq \min(\eps_1^n,\eps_2^n).
\]
If this is not the case, the move is rejected: $(p_1^{n+1},p_2^{n+1},\eps_1^{n+1},\eps_2^{n+1}) = (p_1^n,p_2^n,\eps_1^n,\eps_2^n)$. 
\item if the energy bound is satisfied, compute $a_\dt(v^n,\widetilde{v}^{n+1})$ according to~\eqref{eq:expression_A_dt};
\item generate $U^n \sim \mathcal{U}[0,1]$: if $\log U^n > a_\dt(v^n,\widetilde{v}^{n+1})$, the move is rejected; otherwise it is accepted.
\item if the move is accepted, the new momenta and internal energies are set to
\[
p_1^{n+1} = p_{1}^n + \mu_{12} \left(\widetilde{v}^{n+1} - v^n \right)e_{12}, 
\qquad
p_2^{n+1} = p_{2}^n - \mu_{12} \left(\widetilde{v}^{n+1} - v^n \right)e_{12},
\]
and
\[
\eps_1^{n+1} = \eps_{1}^n - \frac{\mu_{12}}{4} \left[ \left(\widetilde{v}^{n+1}\right)^2 - \left(v^n\right)^2 \right],
\qquad
\eps_2^{n+1} = \eps_{2}^n - \frac{\mu_{12}}{4} \left[ \left(\widetilde{v}^{n+1}\right)^2 - \left(v^n\right)^2 \right].
\]
\end{enumerate}

Let us conclude this section by a quick comment on the computational overhead associated with the Metropolis correction. The first point to mention is that it only concerns the stochastic part of the dynamics. In particular, it does not impact the Hamiltonian part, which is often the most expensive one from a computational viewpoint due to the evaluation of the forces. A second point is that, in order to compute the Metropolis ratio, only two additional terms are needed, namely the ones in the last line of~\eqref{eq:expression_A_dt}; as well as two additional tests (items~(iii) and~(v) in the algorithm above). The overall overhead is therefore quite modest. This should be in any case compensated by a possibly dramatic increase in the timestep for this part of the dynamics, in conjunction with a multiple timestep strategy (see Section~\ref{sec:MTS}).

\subsection{Metropolization of elementary thermal conduction dynamics}
\label{sec:num_TC}

The elementary conduction dynamics between two particles reads
\[
\left\{ \begin{aligned}
d\eps_1 & = \kappa \chi^2(r_{12}) \left(\frac{1}{T_1(\eps_1)}-\frac{1}{T_2(\eps_2)} \right) dt + \sqrt{2\kappa}\chi(r_{12}) \, d\widetilde{W}_{12}, \\
d\eps_2 & = -d\eps_1.
\end{aligned} \right.
\]
Note that it can be rephrased as an effective dynamics on~$\eps_1$ only, upon introducing $E_{12,0} = \eps_{1,0} + \eps_{2,0}$:
\[
d\eps_1 = \kappa \chi^2(r_{12}) \Big( s_1'(\eps_1) - s_2'(E_{12,0} - \eps_1) \Big) dt + \sqrt{2\kappa}\chi(r_{12}) \, d\widetilde{W}_{12}.
\]
The latter dynamics is a stochastic differential equation of overdamped Langevin type, with invariant probability measure $Z^{-1} \, \rme^{s_1(\eps_1) + s_2(E_{12,0}-\eps_1)} \, d\eps_1$. It is therefore possible to find accurate schemes for this dynamics, and to stabilize them with a Metropolis correction by following the same approach as in Section~\ref{sec:Metropolis_elementary_FD}.

More precisely, consider the following numerical scheme: starting from two energies $\eps_1^n,\eps_2^n$ for which $\chi(r_{12})>0$ (otherwise nothing needs to be done), compute the energy increment
\[
\Delta \eps^n = \kappa \dt \chi^2(r_{12}) \Big( s_1'(\eps_1^n) - s_2'(\eps_2^n) \Big) + \sqrt{2\kappa\dt }\chi(r_{12}) \, \widetilde{G}^n,
\]
and propose 
\[
\widetilde{\eps}_1^{n+1} = \eps_1^n + \Delta \eps^n, \qquad \widetilde{\eps}_2^{n+1} = \eps_2^n - \Delta \eps^n.
\]
If either $\widetilde{\eps}_1^{n+1} \leq 0$ or $\widetilde{\eps}_2^{n+1} \leq 0$, then the move is rejected and $(\eps_1^{n+1},\eps_2^{n+1}) = (\eps_1^{n},\eps_2^{n})$. Otherwise, the move is accepted with probability $\min\left(1,\rme^{a_\dt(\eps_1^n,\eps_2^n,G^n)}\right)$,
where 
\[
a_\dt(\eps_1^n,\eps_2^n,G^n) = s_1(\widetilde{\eps}_1^{n+1})+s_2(\widetilde{\eps}_2^{n+1})-\Big( s_1(\eps_1^n)+s_2(\eps_2^{n}) \Big) + \frac12 \left( \left|\widetilde{G}^n\right|^2 - \left|\widehat{G}^n\right|^2 \right),
\]
with
\[
\widehat{G}^n = \frac{1}{\sqrt{2\kappa\dt}\chi(r_{12})}\left( \eps_1^n - \widetilde{\eps}_1^{n+1} - \kappa \dt \chi^2(r_{12}) \Big( s_1'(\widetilde{\eps}_1^{n+1}) - s_2'(\widetilde{\eps}_2^{n+1}) \Big) \right).
\]

\subsection{Error estimates on thermodynamic averages}
\label{sec:errors}

The numerical schemes $\Phi_\dt^{{\rm FD},ij}$ and $\Phi_\dt^{{\rm TC},ij}$ presented in Sections~\ref{sec:Metropolis_elementary_FD} and~\ref{sec:num_TC} preserve by construction the measure~\eqref{eq:nu_N} because of the Metropolis correction. On the other hand, the Verlet scheme is only second-order accurate. Using an analysis similar to the one performed in~\cite{LMS16} for Langevin dynamics, and under appropriate ergodicity assumptions (both for the continuous dynamics and its discrete approximation), it can then be shown that, for any physical observable~$\varphi$,
\begin{equation}
  \label{eq:bias_inv_meas}
  \int \varphi \, d\nu_{N,\dt} = \int \varphi \, d\nu_{N} + \mathrm{O}(\dt^2),
\end{equation}
where $\nu_{N,\dt}$ is the probability measure which is actually sampled by the numerical scheme with a timestep~$\dt$. This equality means that average properties, as obtained for instance by time averages over a very long numerical DPDE trajectory, coincide with the thermodynamic averages with respect to the measure~\eqref{eq:nu_N} up to a systematic error of order~$\dt^2$. Of course, the above reasoning is only formal since ergodicity cannot be proved in general for DPD-like systems -- the only known result is for one-dimensional DPD~\cite{SY06}.

In practice, the total energy~$\cE$ drifts in time when DPDE is discretized unless some projection procedure is enforced (for instance by rescaling internal energies in order to keep the total energy constant; see~\cite{LBBA11,HMS16} for discussions on this issue). When no energy rescaling is used, as is the case for some of the simulations reported in Section~\ref{sec:numerics}, ergodic (infinite time) averages cannot be considered. Finite time averages are the only quantities which make sense. 

\section{Numerical illustrations}
\label{sec:numerics}

The numerical schemes considered in this section are given by~\eqref{eq:splitting_scheme}, except in the second part of Section~\ref{sec:LJ}. The system is spatially decomposed using a linked-cell method, so the order of integration of the various pairs may change from one step to the other. Let us emphasize that, for all simulations reported below, even for the smallest timesteps and in the absence of potential energy functions, standard SSA simulations (\text{i.e.} without Metropolis correction) always crashed after a short time due to the appearance of negative internal energies; so that no numerical result could be reported in those cases. This is however possible for heat capacities larger than the ones considered here, and/or a smaller Einstein temperature $\To$. In such situations, the biases/systematic errors related to the timestep are of the same order of magnitude irrespectively of the fact that the elementary dynamics are Metropolized or not. This highlights the fact that the Metropolis correction is really useful to stabilize the dynamics rather than to reduce a possibly large bias.

Unless otherwise mentioned, the system under consideration is composed of $N = 1600$ particles in dimension $d=2$, at particle density $\rho = 1$, with pairwise interactions:
\[
V(q) = \sum_{1 \leq i < j \leq N} u(|q_i-q_j|).
\]
The temperature is set to $\Tinit = 1$ for initialization (see Section~\ref{sec:initialization}), the fluctuation magnitude is chosen to be $\sigma^2 = 2$ (in fact, $\sigma^2 = 2\gamma_* \Tinit$ with $\gamma_* = 1$), and the thermal conductivity is set to $\kappa = 1$. The cut-off radius is $r_{\rm cut} = 3$ for the fluctuation/dissipation and thermal interactions, with weight function
\[
\chi(r) = \left\{ \begin{aligned}
1 - \frac{r}{r_{\rm cut}} & \qquad \text{for } r \leq r_{\rm cut}, \\
0 & \qquad \text{for } r \geq r_{\rm cut}.
\end{aligned} \right.
\]
The micro-EOS used in all simulations below is the blended Einstein model~\eqref{eq:s_blended} with $\Cv = 5$, $\Cvo = 1$ and $\To = 1$ (the same parameters as in Figure~\ref{fig:Einstein}). Reduced units where $k_{\rm B} = 1$, $\epsref = 1$ and $m=1$ are used throughout. Average properties are estimated by time averages over a simulation time $\tau_{\rm simu} = 10^4$.

\subsection{Creation of initial conditions and thermalization}
\label{sec:initialization}

The system starts from a solid phase with atoms on a cubic lattice, and velocities sampled according to the Boltzmann distribution at temperature $\Tinit$ (whose associated inverse temperature is denoted by $\betainit$). The system is next integrated for a time $\tau_{\rm therm}$ with a timestep $\dt$ using a Langevin dynamics at friction $\gamma_*$ (using the so-called Geometric Langevin Algorithm introduced in~\cite{BO10} and also studied in~\cite{LMS16}). Internal energies are sampled independently from $Z_\eps^{-1} \exp(s(\eps)-\betainit \eps) \, d\varepsilon$, by discretizing the one-dimensional overdamped Langevin dynamics for each internal energy~$\eps_i$, as done in~\cite{Stoltz06}:
\[
d\eps_t = - \left(1 - \frac{s'(\eps)}{\betainit}\right) dt + \sqrt{\frac{2}{\betainit}} \, dW_t = - \left(1 - \frac{\Tinit}{T(\eps)}\right) dt + \sqrt{\frac{2}{\betainit}} \, dW_t.
\]
In practice, this dynamics is discretized with a Euler-Maruyama scheme and an effective timestep $\dt_{\rm eff} = \Cv \dt$ as
\[
\widetilde{\eps}^{n+1} = \eps^n - \left(1 - \frac{s'(\eps^n)}{\betainit}\right) \dt_{\rm eff} + \sqrt{\frac{2\dt_{\rm eff}}{\betainit}} \, G^n,
\]
and corrected by a Metropolis procedure: the proposal $\widetilde{\eps}^{n+1}$ is accepted with probability $\min(1,\rme^{a_\dt(\eps^n,G^n)})$ where
\[
a_\dt(\eps^n,G^n) = s\left(\widetilde{\eps}^{n+1}\right) - s(\eps^n) - \betainit \left(\widetilde{\eps}^{n+1} - \eps^n\right) + \frac12 \left(|G^n|^2-\left|\widehat{G}^n\right|^2\right),
\]
with 
\[
\widehat{G}^n = \sqrt{\frac{\betainit}{2\dt_{\rm eff}}}\left(\eps^n - \widetilde{\eps}^{n+1} +  \left(1 - \frac{s'(\widetilde{\eps}^{n+1})}{\betainit}\right) \dt_{\rm eff} \right).
\]

The thermalization time is set to $\tau_{\rm therm} = 20$. At the end of the thermalization, one typical configuration sampled according to the canonical measure~\eqref{eq:canonical_N} is obtained (with a small bias due to the timestep errors). An additional burn-in is performed for a time $\tau_{\rm burn-in} = 20$, using DPDE. The timestep for these integrations is $\dt = 0.01$ in all cases, except for Lennard--Jones systems where it is set to $\dt = 0.001$. Let us remark that, at the end of this equilibration, a typical configuration for the measure~\eqref{eq:nu_N} is obtained. However, since the equivalence of ensembles between~\eqref{eq:nu_N} and~\eqref{eq:canonical_N} holds only in the limit $N \to +\infty$, there is a priori a bias on thermodynamic properties between the averages with respect to these two measures (which should be of order~$1/N$). An additional bias arises from the finiteness of the timestep used in the equilibration. These biases explain why the average temperatures computed in the simulations reported in Section~\ref{sec:biases} converge to a value close to, but different from $\Tinit$ when the timestep of the simulation converges to~0.

\subsection{Timestep biases for various systems}
\label{sec:biases}

As already mentioned in Section~\ref{sec:errors}, only errors on finite time averages are considered since the energy may drift in time (the drift rate increasing with $\dt$). Error bars on finite time averages are in all cases of the order of a few percents at most, and are hence omitted. 

The aim of the simulation results reported below is first and foremost to demonstrate that quite large timesteps can be used to integrate the dynamics. Average energy drifts are however reported for some representative choices of parameters (no picture is provided since the phenomenon has been described at length in~\cite{HMS16}). These drifts are obtained by performing several independent realizations of the dynamics for a given initial condition, and computing the average energy over the various realizations as a function of time. As in previous studies (see~\cite{LBBA11,HMS16}), the systematic drift is observed to be linear in time and quite small for timesteps which are not too large. 

For larger timesteps and when thermodynamic (infinite time) averages are of interest, the numerical scheme proposed have to be complemented by some projection procedure to enforce the energy conservation~\cite{LBBA11,HMS16}. More precisely, given a total energy $\mathcal{E}^n$ at step~$n$ and a new configuration $(q^{n+1},p^{n+1},\eps^{n+1})$ obtained after one step of the splitting algorithm (with possibly several substeps of the Hamiltonian part when multiple timestep strategies are used), the internal energies are rescaled by a factor
\begin{equation}
  \label{eq:rescaling_energy}
  \alpha^{n+1} = \frac{\mathcal{E}^n - H(q^{n+1},p^{n+1})}{\sum_{i=1}^N \eps_i^{n+1}}.
\end{equation}
Note that, by construction, $\mathcal{E}(q^{n+1},p^{n+1},\alpha^{n+1}\eps^{n+1}) = \mathcal{E}^n$. Let us emphasize that this projection does not change the stability properties of the algorithm: a new configuration $(q^{n+1},p^{n+1},\eps^{n+1})$ obtained by one step of the integration scheme is needed in any case. The projection does not allow for larger timesteps; it only avoids energy drifts in the long term.

\subsubsection{Ideal fluid}

\begin{figure}
\begin{center}
\includegraphics[width=0.5\textwidth]{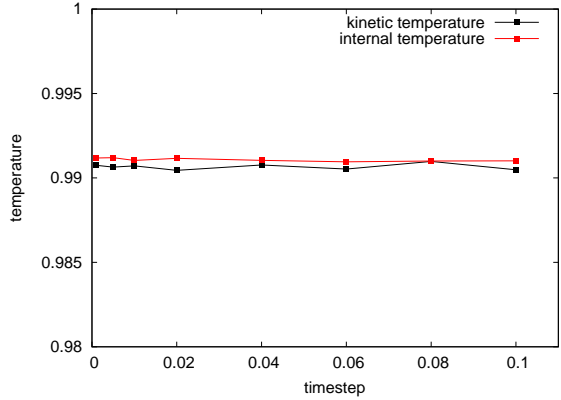}
\end{center}
\caption{Kinetic and internal temperatures as a function of the timestep for the ideal fluid.}
\label{fig:ideal}
\end{figure}

Consider first ideal fluids, which correspond to the trivial interaction potential $u(r)=0$; see Figure~\ref{fig:ideal}. The first element to note is that there is no timestep restriction for the Metropolized scheme, and that there is no bias, even for very large timesteps. There are also no energy drifts since energy is exactly preserved. Yet, the rejection rate is very small: for the largest timestep ($\dt = 0.1$), it is below $10^{-3}$, while it is of order $5 \times 10^{-6}$ for $\dt = 0.001$. These rare rejections are however crucial in ensuring the stability of the dynamics. In particular, the number of counts for proposed negative energies is of the order of $10^5$ for all simulations.

\subsubsection{Soft fluid}

\begin{figure}
\includegraphics[width=0.5\textwidth]{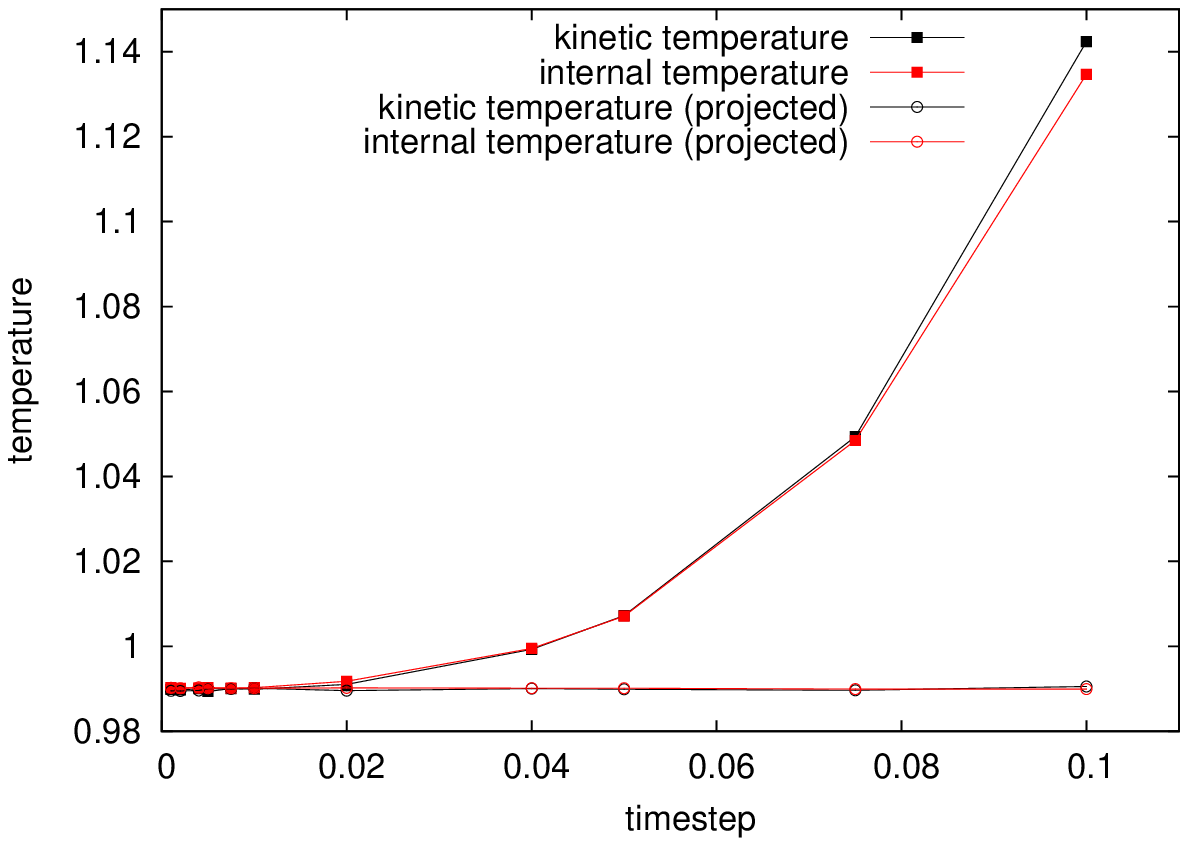}
\includegraphics[width=0.5\textwidth]{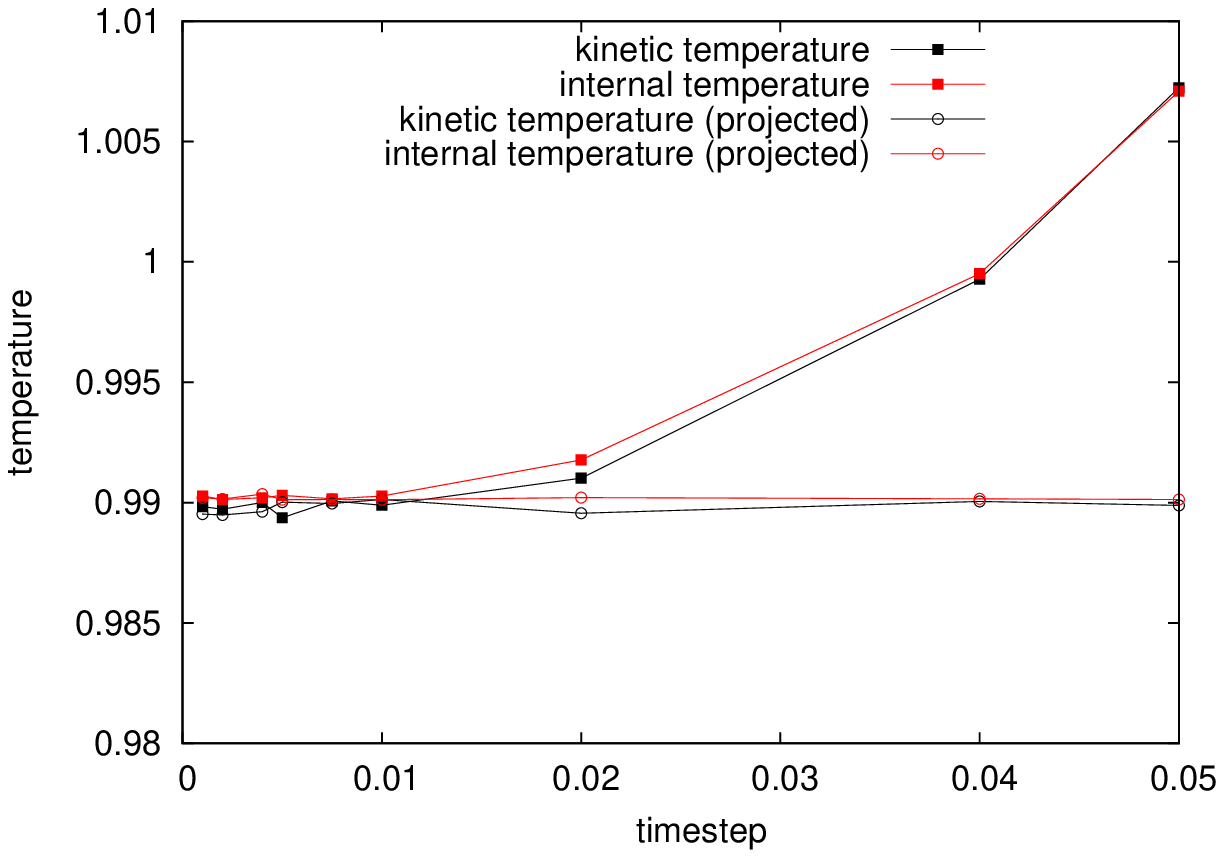}
\caption{Kinetic and internal temperatures as a function of the timestep for the soft DPD fluid, with and without the energy projection encoded by~\eqref{eq:rescaling_energy}. The right picture provides a zoom on the small timestep values.}
\label{fig:soft}
\end{figure}

Consider next a soft interaction potential of the form
\[
u(r) = \left\{ \begin{aligned}
& \varepsilon_{\rm DPD}\left(1 - \frac{r}{r_{\rm cut}}\right)^2 & \text{for } r \leq r_{\rm cut}, \\
& 0 & \text{for } r \geq r_{\rm cut}.
\end{aligned} \right.
\]
The results are presented in reduced units where the reference energy corresponds to $\varepsilon_{\rm DPD}=1$; see Figure~\ref{fig:soft}. Here again, it is seen that the Metropolized scheme is unconditionnally stable (\textit{i.e.} any simulation timestep can be considered). The average energy drift, estimated by the procedure described in~\cite{HMS16}, is linear in time. The relative increase in energy is of order $10^{-8}$ per unit time for $\dt = 0.01$, but increases to $2 \times 10^{-5}$ for $\dt = 0.1$. The rejection rates are comparable to the ones observed for the ideal fluid. Note also that the bias starts off quadratically when no projection is used (as would be predicted by~\eqref{eq:bias_inv_meas} for the Metropolized scheme if the dynamics was ergodic; which is not the case here since the energy drifts in time). With the energy projection procedure encoded by~\eqref{eq:rescaling_energy}, there is almost no bias, as already observed in~\cite{LBBA11,HMS16}. Let us therefore emphasize again that the main interest of the approach we describe in this work lies in the increased stability properties of the method: with the small heat capacities we consider, it is not possible to perform simulation without correcting for negative energies.

\subsubsection{Lennard-Jones fluid}
\label{sec:LJ}

Consider finally the splined Lennard--Jones potential
\[
u(r) = \left\{ 
\begin{aligned}
&4\varepsilon_{\rm LJ}\left[\left(\frac{\sigma_{\rm LJ}}{r}\right)^{12}-\left(\frac{\sigma_{\rm LJ}}{r}\right)^{6}\right] & \text{for } r \leq r_{\rm spline}, \\
&(A + Br) (r-r_{\rm cut})^2 & \text{for } r_{\rm spline} \leq r \leq r_{\rm cut}, \\
&0 & \text{for } r \leq r_{\rm cut},
\end{aligned} \right.
\]
with $r_{\rm spline} = \delta r_{\rm cut}$, and where $A,B$ are chosen in order to ensure that $u$ and its first derivative are continuous. The simulations are performed in reduced units, with $\delta = 0.8$ and $\varepsilon_{\rm LJ} = 1$, $\sigma_{\rm LJ}=1$. The relative rate of increase of the total energy per unit time is again quite small, of order $10^{-5}$ for $\dt = 0.005$; and negligible (below $10^{-8}$) for multiple timestep strategies with a timestep of $0.001$ for the Hamiltonian part. 

\begin{figure}
\begin{center}
\includegraphics[width=0.5\textwidth]{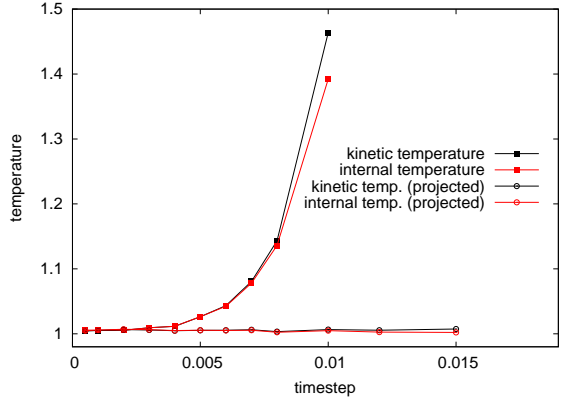}
\end{center}
\caption{Kinetic and internal temperatures as a function of the timestep for the LJ fluid (with and without projection to enforce total energy conservation).}
\label{fig:LJ}
\end{figure}

The results for the biases are reported in Figure~\ref{fig:LJ}. Note that, below the stability treshold of the method, around $\dt = 0.005$, there is almost no bias. The stability is in fact limited by the singularities of the Lennard-Jones potential, as made clear when resorting to the multiple-timestep strategy. When a projection is used, the stability is even slightly better because higher energy states, which require even smaller timesteps for the integration, are not visited since there is no drift in the energy; so that larger timesteps can be considered.

\begin{figure}
\begin{center}
\includegraphics[width=0.49\textwidth]{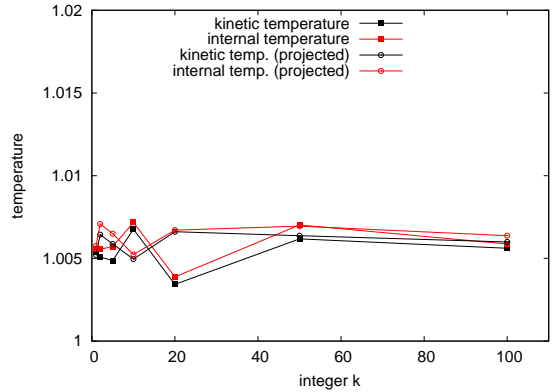}
\end{center}
\caption{Average kinetic and internal temperatures as a function of the timestep, for a Lennard--Jones system and various values of $k_{\rm MTS}$, with $\dt_{\rm Ham} = 0.001$ fixed (with and without projection to enforce total energy conservation).}
\label{fig:MTS}
\end{figure}

Figure~\ref{fig:MTS} presents simulation results obtained using the multiple timestep scheme~\eqref{eq:splitting_scheme_MTS}, with $\dt_{\rm Ham} = 0.001$, and various values of the integer $k_{\rm MTS}$. The bias is almost constant with increasing $k_{\rm MTS}$, which shows that the errors on the invariant measure really arise from the Hamiltonian part of the dynamics. The energy drift is in fact very small, even for $k_{\rm MTS} = 100$ which corresponds to a timestep $\Delta t = 0.1$ for the stochastic parts of the dynamics. The total energy projection therefore has no noticeable impact on the results in this case.

\subsection{Equilibration dynamics}
\label{sec:equilibration}

The final illustration is the simulation of a transient relaxation, where the initial condition is obtained by equilibrating internal energies at a given temperature $T_{{\rm int},0}$, while the mechanical degrees of freedom~$(q,p)$ are sampled at a temperature $T_{{\rm mech},0}$. The initialization is performed as described in Section~\ref{sec:initialization}, except that the temperatures are different for internal and mechanical degrees of freedom. 

\begin{figure}
\begin{center}
\includegraphics[width=0.49\textwidth]{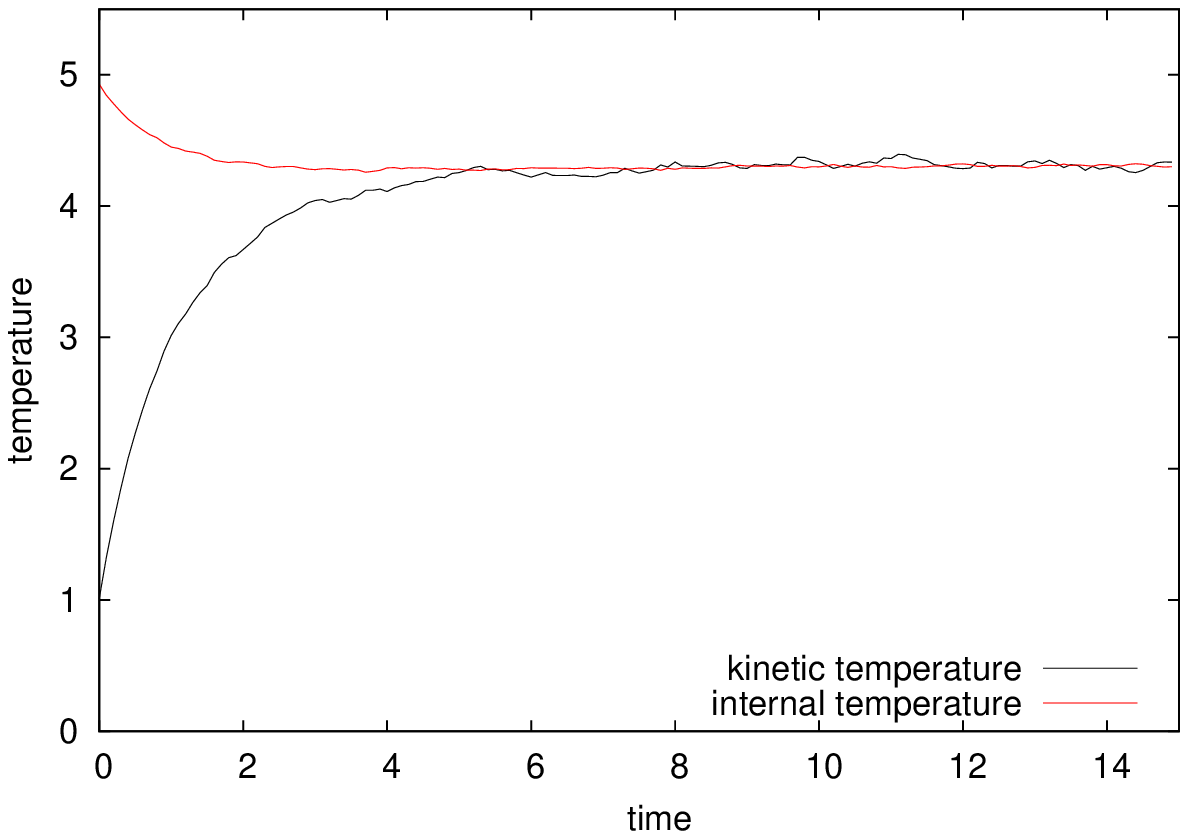}
\includegraphics[width=0.49\textwidth]{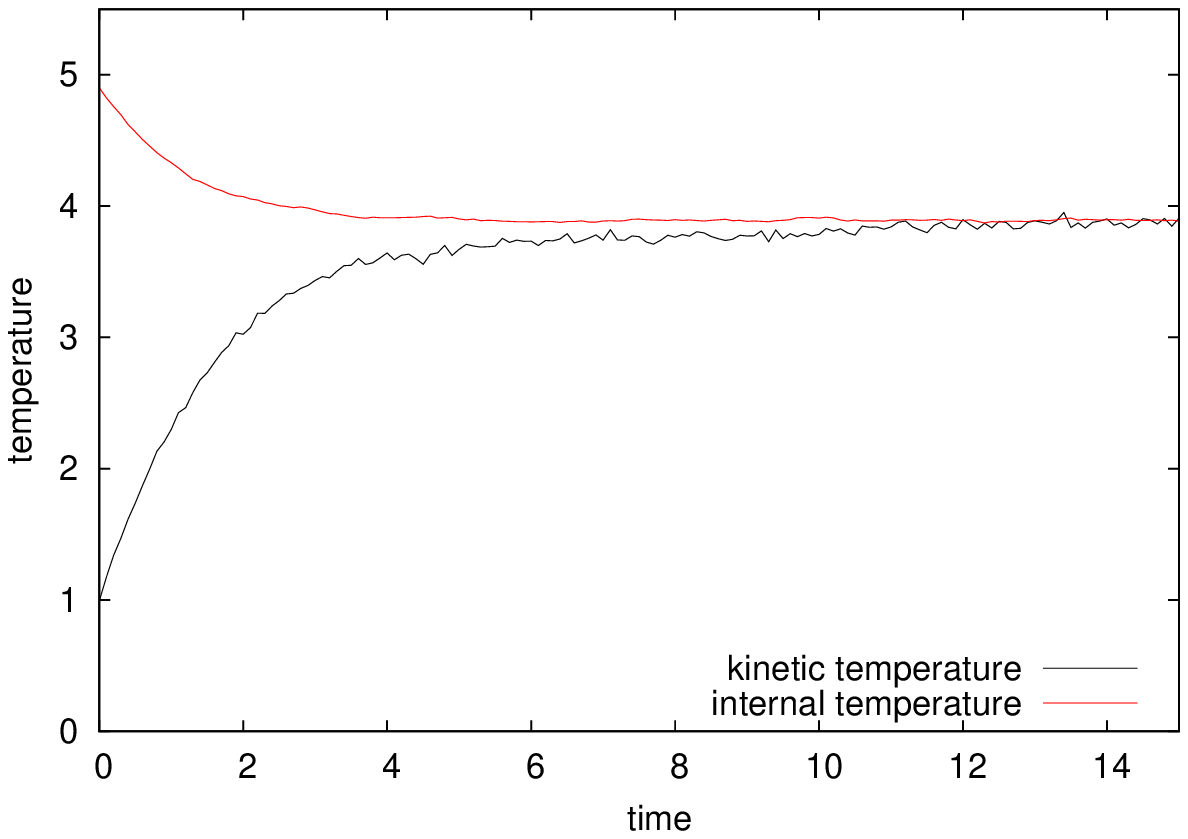}
\end{center}
\caption{Kinetic and internal temperatures as a function of the time for equilibration dynamics. Left: Soft DPD potential. Right: Lennard--Jones potential.}
\label{fig:equilibration}
\end{figure}

The system under study is larger than in the previous section, namely $N=10^4$ particles in a 2D setting, the other parameters being unchanged. The initial termalization time is set to $\tau_{\rm therm} = 100$, with $T_{{\rm int},0}=5$ and $T_{{\rm mech},0}=1$. After the time $\tau_{\rm therm}$, the internal and kinetic temperatures are monitored; see Figure~\ref{fig:equilibration} ($\dt = 0.01$ for the soft DPD potential, while $\dt = 0.001$ for Lennard--Jones systems). It is expected that they converge to a common value after a certain physical time, which is mostly dictated by the fluctuation magnitude $\sigma$. Such equilibration dynamics are used to parameterize the fluctuation/dissipation in DPDE~\cite{KSM16}. Another option, considered in~\cite{LBBA11}, consists in instantaneously heating only a part of the system, in which case the thermal conduction has a stronger influence.

\section{Conclusion and perspectives}
\label{sec:ccl}

This article introduced new integration schemes for DPDE, using a splitting paradigm similar to SSA, but based on the integration of elementary pairwise fluctuation/dissipation and thermal conduction dynamics seen as effective dynamics of a single variable. The interest of such reformulations is that the numerical schemes for the elementary dynamics can be corrected by a Metropolis procedure, which dramatically improves the numerical stability of the algorithm (preventing by construction the occurence of negative internal energies, even for arbitrarily small heat capacities); and also leads to biases which are small. The increased stability properties of the stochastic part furthermore suggest to resort to multiple timestep strategies, where the Hamiltonian part is integrated with a small timestep (dictated by stability conditions) while the stochastic part is integrated less frequently but with larger timesteps. Such approaches are very interesting from a computational viewpoint since the stochastic part of the dynamics can computationally quite expensive. Of course, the schemes presented here can be combined in actual applications with some projection procedure, which improves the computation of average properties at equilibrium but is dubious for nonequilibrium systems (see the discussion in~\cite{HMS16}).

The approach outlined here for DPDE can of course be straightforwardly implemented for standard DPD. It can also be extended to smoothed dissipative particle dynamics when the latter dynamics is reformulated in terms of internal energies rather than internal entropies~\cite{FRMS16}.

Our focus here mostly was on thermodynamic (static) averages. An interesting question, only hinted at with the numerical results on the transient equilibration in Section~\ref{sec:equilibration}, and left aside for a subsequent work, is the dynamical relevance of the algorithm presented here -- for instance for the computation of transport coefficients or the simulation of nonequilibrium shock and detonation waves.

Finally, another valuable line of research is to adapt the method presented here so that it can be used in the current massively parallel implementations of DPDE~\cite{LBMLM14,HMS16}.

\bigskip

\paragraph{Acknowledgements}
G. Stoltz acknowledges stimulating discussions with John Brennan, as well as a longstanding collaboration with Jean-Bernard Maillet. These two researchers both highlighted several practical issues in the integration of DPDE, in particular the occurence of negative internal energies. This work benefited from a funding of the European Army Research Office, under grant award \#W911NF-16-1-0254. It is also supported by the European Research Council under the European Union's Seventh Framework Programme (FP/2007-2013) / ERC Grant Agreement number 614492; and by the Agence Nationale de la Recherche, under grant ANR-14-CE23-0012 (COSMOS).

\section*{Appendix~A: Properties of the Einstein micro-EOS}
\label{sec:properties_Einstein_EOS}

It is shown in this appendix that the micro-EOS~\eqref{eq:s_Einstein} leads to the heat capacity predicted by the Einstein model of harmonic oscillators. The temperature $\To$ allows to define the scale of energies for which quantum effects are non negligible. In the limit $\To \to 0$, the classical micro-EOS 
\[
s(\eps) = \frac{\Cv}{k_{\rm B}} \left[1+\ln \left(\frac{\eps}{\eps_{\rm ref}}\right)\right].
\]
is recovered (compare with~\eqref{eq:s_classical}; the extra additive constant $\Cv/k_{\rm B}$ is unimportant).

Note first, that, from the definition of the internal entropy~\eqref{eq:s_Einstein},
\begin{equation}
\label{eq:derivative_microEOS}
s'(\eps) = -\frac{1}{\kB \To}\ln\left(\frac{\eps}{\eps + \Cv\To}\right).
\end{equation}
In view of~\eqref{eq:def_temp_microEOS}, it follows that
\begin{equation}
  \label{eq:temp_fct_eps}
  T(\eps) = -\frac{\To}{\dps \ln\left(1 - \frac{\Cv \To}{\eps + \Cv\To}\right)}.
\end{equation}
On this expression, it is clear that $T(\eps) \sim \eps/\Cv$ as $\eps \to +\infty$ for $\To$ fixed, and that $T(\eps) \to \eps/\Cv$ as $\To \to 0$ for $\eps > 0$ fixed. Both limits are a signature that quantum effects are negligible.

Let us now make explicit the model heat capacity which underlines the model~\eqref{eq:s_Einstein}, in order to motivate that the small energy behavior is appropriate. From~\eqref{eq:eps_integral_Cv}, it follows that, for any microEOS, 
\begin{equation}
\label{eq:Cv_parametric_curve}
C_v(T(\eps)) = \frac{1}{T'(\eps)}.
\end{equation}
On the other hand, the relation~\eqref{eq:temp_fct_eps} can be inverted to write the energy as a function of the temperature. More precisely, 
\begin{equation}
  \label{eq:reformulation_T_eps}
  1 - \frac{\Cv \To}{\eps + \Cv\To} = \rme^{-\To/T(\eps)} 
\end{equation}
so that 
\[
\left(1 - \rme^{-\To/T(\eps)} \right) \eps = \Cv\To \rme^{-\To/T(\eps)},
\]
and finally, the energy $\eps$ can be written in terms of the temperature $\theta$ as
\[
\eps(\theta) = \frac{\Cv\To \rme^{-\To/\theta}}{1 - \rme^{-\To/\theta}}.
\]
Since, using~\eqref{eq:reformulation_T_eps},
\[
\rme^{-\To/T(\eps)} \frac{\To T'(\eps)}{T(\eps)^2} = \frac{\Cv \To}{(\eps + \Cv\To)^2},
\]
it follows that
\[
T'(\eps) = \frac{\Cv T(\eps)^2}{(\eps + \Cv\To)^2} \rme^{\To/T(\eps)}, 
\]
This leads therefore to 
\begin{equation}
  \label{eq:Cv_theta_Einstein}
  C_v(\theta) = \frac{1}{T'(\eps(\theta))} = \frac{(\eps(\theta) + \Cv\To)^2}{\Cv \theta^2} \rme^{-\To/\theta} = \Cv \, \left(\frac{\To}{\theta}\right)^2 \frac{\rme^{-\To/\theta}}{\left(1 - \rme^{-\To/\theta}\right)^2}.
\end{equation}
It is easy to check that $C_v(\theta) \sim \Cv$ when $\theta \to +\infty$, while $C_v(\theta) \sim (\To/\theta)^{-2} \, \rme^{-\To/\theta}$ vanishes at all orders as $\theta \to 0$. The model heat capacity considered is the one corresponding to the Einstein model of harmonic oscillators.

\subsection*{Estimators of the thermodynamic temperature from the internal energies}

Recall that the marginal of the canonical measure~\eqref{eq:canonical_N} in the variable~$\eps_i$ is given by~\eqref{eq:marginal_canonical_eps}. When the internal entropies are such that 
\begin{equation}
\label{eq:conditions_s}
\forall i=1,\dots,N, \qquad s_i(\eps_i) \xrightarrow[\eps_i \to 0]{} -\infty, \qquad s_i(\eps_i)-\beta \eps_i \xrightarrow[\eps_i\to+\infty]{} -\infty,
\end{equation}
an integration by parts shows that
\begin{equation}
\label{eq:std_int_temp}
\begin{aligned}
\left\langle \frac{1}{\kB T_i(\eps_i)} \right\rangle_{\mu_\beta} & = \frac{\dps \int_0^{+\infty} s_i'(\eps_i) \, \rme^{s_i(\eps_i)-\beta \eps_i}\, d\eps_i}{\dps \int_0^{+\infty} \rme^{s_i(\eps_i)-\beta \eps_i}\, d\eps_i} = \beta + \frac{\dps \int_0^{+\infty} \left( s_i'(\eps_i) - \beta \right) \rme^{s_i(\eps_i)-\beta \eps_i}\, d\eps_i}{\dps \int_0^{+\infty} \rme^{s_i(\eps_i)-\beta \eps_i}\, d\eps_i} \\
& = \beta + \frac{\left[\rme^{s_i(\eps_i)-\beta \eps_i}\right]_0^{+\infty}}{\dps \int_0^{+\infty} \rme^{s_i(\eps_i)-\beta \eps_i}\, d\eps_i} = \beta.
\end{aligned}
\end{equation}
This motivates taking harmonic averages of the internal temperatures as an estimator of the thermodynamic temperature for micro-EOS satisfying~\eqref{eq:conditions_s} (such as~\eqref{eq:s_classical} and~\eqref{eq:s_blended}). 

On the other hand, the internal entropy~\eqref{eq:s_Einstein} obtained from the Einstein model of harmonic oscillators is not such that $s(\eps) \to -\infty$ as $\eps \to 0$. Alternative estimators of the internal temperature are therefore required. For a general function $F \in C^1$ such that 
\[
F(0) = 0, \qquad \lim_{\eps_i \to +\infty} F(\eps_i) \, \rme^{s_i(\eps_i)-\beta \eps_i} = 0,
\]
an integration by parts similar to the one used above shows that
\[
\begin{aligned}
\int_0^{+\infty} F(\eps_i) \, s_i'(\eps_i) \, \rme^{s_i(\eps_i)-\beta \eps_i}\, d\eps_i & = \beta \int_0^{+\infty} F(\eps_i) \, \rme^{s_i(\eps_i)-\beta \eps_i}\, d\eps_i - \int_0^{+\infty} F'(\eps_i) \, \rme^{s_i(\eps_i)-\beta \eps_i}\, d\eps_i \\
& \ \ + \left[F(\eps_i) \rme^{s_i(\eps_i)-\beta \eps_i}\right]_0^{+\infty},
\end{aligned}
\]
from which the following estimator of the thermodynamic temperature is deduced:
\begin{equation}
  \label{eq:estimator_temperature_Einstein}
  \frac{\left\langle F(\eps_i)\right\rangle_{\mu_{\rm int}}}{\left\langle F(\eps_i)\,s_i'(\eps_i) + F'(\eps_i)\right\rangle_{\mu_{\rm int}}} = \frac1\beta,
\end{equation}
where $\mu_{\rm int}$ is defined in~\eqref{eq:marginal_canonical_eps}. Note that this estimator is a ratio of canonical averages (similar to what is considered to estimate the potential temperature using the Laplacian and the gradient of the potential~\cite{BAJE98}). One possible choice for $F$ is $F(\eps) = \eps$, which leads to the estimator~\eqref{eq:estimator_temperature_eps}. More generally, higher order moments of the internal energy can be used by considering $F(\eps) = \eps^n$ for $n \geq 1$.

\section*{Appendix~B: Generalized fluctuation-dissipation}
\label{sec:generalized_FD}

It is shown in this appendix how to extend the derivation of Sections~\ref{sec:rewriting} and~\ref{sec:Metropolis_elementary_FD} to anisotropic fluctuation/dissipation dynamics, with components both along lines of centers and orthogonal to this direction, and possibly of different magnitudes. The most general dynamics is first presented, and then specified to the case when the fluctuation/dissipation can be decomposed into parallel and orthogonal components as in~\cite{JPK08}. It is finally explained how to implement the Metropolis correction.

\paragraph{General dynamics} The elementary dynamics~\eqref{eq:DPDE_elementary_12} on the momenta can be generalized as 
\begin{equation}
\label{eq:DPDE_elementary_anisotropic}
\left \{
\begin{aligned}
dp_1 & = -\bgamma(\br_{12},\eps_1,\eps_2) v_{12} \, dt + \bsigma(\br_{12}) \, dW_t, \\
dp_2 & = -dp_1,
\end{aligned}
\right.
\end{equation}
where $\br_{12} = q_1-q_2$ is a $d$-dimensional vector (with $d$ the underlying physical dimension), $W_t$ is a standard $d$-dimensional Brownian motion, and $\bgamma(r),\bsigma(r)$ are functions with values in the space of $d \times d$ real matrices. The evolution of the internal energies is deduced from the conservation of the elementary kinetic plus internal energies. Using It\^o calculus, 
\[
d\eps_1 = d\eps_2 = \frac12 \left[v_{12}^T \bgamma(\br_{12},\eps_1,\eps_2) v_{12} - \frac{1}{2\mu_{12}} \mathrm{Tr}\left(\bsigma\bsigma^T\right)(\br_{12}) \right] - \frac12 v_{12}^T \bsigma(\br_{12}) dW_t.
\]
The measure~\eqref{eq:nu_N} is invariant provided
\[
\bgamma(\br_{12},\eps_1,\eps_2) = \frac{1}{4k_{\rm B}} \left(\frac{1}{T_1(\eps_1)}+\frac{1}{T_2(\eps_2)}\right)\bsigma(\br_{12})\bsigma(\br_{12})^T.
\]

A simple computation shows that the variations of the kinetic energy can be fully understood in terms of the variations of the relative velocity. More precisely, \eqref{eq:kin_energy_difference} and~\eqref{eq:eps_in_terms_of_v} should be replaced with
\[
\frac{p_1^2 - p_{1,0}^2}{2m_1} + \frac{p_2^2 - p_{2,0}^2}{2m_2} = \frac{\mu_{12}}{2} \left[ \left(v_{12}\right)^2 - \left(v_{12,0}\right)^2 \right]. 
\]
and
\[
\eps_1 = \eps_{1,0} - \frac{\mu_{12}}{4} \left[ \left(v_{12}\right)^2 - \left(v_{12,0}\right)^2 \right], \qquad
\eps_2 = \eps_{2,0} - \frac{\mu_{12}}{4} \left[ \left(v_{12}\right)^2 - \left(v_{12,0}\right)^2 \right]. 
\]

\paragraph{Parallel and orthogonal fluctuation/dissipation} One can typically consider scalar friction and fluctuation coefficients $\gamma^\pr, \gamma^\perp,\sigma^\pr, \sigma^\perp$, which depend on whether the friction and fluctuation are parallel to the lines of centers or orthogonal to this direction, as well as associated othogonal projection matrices $P^\pr(\br), P^\perp(\br) \in \mathbb{R}^{d \times d}$ and cut-off functions~$\chi^\pr(r),\chi^\perp(r)$ (depending on $r = |\br|$). In this case,
\begin{equation}
\label{eq:general_friction_fluctuation}
\begin{aligned}
\bgamma(\br_{12},\eps_1,\eps_2) & = \gamma^{\pr}(\eps_1,\eps_2) \chi^\pr(r)^2 P^\pr(\br) + \gamma^{\perp}(\eps_1,\eps_2) \chi^\perp(r)^2 P^\perp(\br), \\
\bsigma(r) & = \sigma^{\pr} \chi^\pr(r) P^\pr(\br) + \sigma^{\perp} \chi^\perp(r) P^\perp(\br),
\end{aligned}
\end{equation}
where 
\[
P^\pr(\br) = \frac{\br}{r} \otimes \frac{\br}{r}, 
\qquad 
P^\perp(\br) = \mathrm{Id} - P^\pr(\br).
\]
The invariance of the measure~\eqref{eq:nu_N} is then a consequence of the following standard scalar conditions on each component (similar to~\eqref{eq:FDR1}):
\[
\gamma^\pr(\eps_1,\eps_2) = \frac{\left(\sigma^\pr\right)^2}{4k_{\rm B}} \left(\frac{1}{T_1(\eps_1)}+\frac{1}{T_2(\eps_2)}\right),
\qquad
\gamma^\perp(\eps_1,\eps_2) = \frac{\left(\sigma^\perp\right)^2}{4k_{\rm B}} \left(\frac{1}{T_1(\eps_1)}+\frac{1}{T_2(\eps_2)}\right).
\] 

The dynamics~\eqref{eq:DPDE_elementary_anisotropic} can be rewritten as
\[
dv_{12} = - \frac12 \bB(\br_{12})^2 \nabla U(v_{12})\,dt + \bB(r_{12}) \, dW_t,
\]
with
\[
\bB(r_{12}) = \frac{\bsigma(\br_{12})}{\mu_{12}},
\]
and 
\[
U(v) = -s_1\left(\eps_{1,0} - \frac{\mu_{12}}{4} \left[ v^2 - \left(v_{12,0}\right)^2 \right]\right) - s_2\left(\eps_{2,0} - \frac{\mu_{12}}{4} \left[ v^2 - \left(v_{12,0}\right)^2 \right]\right).
\]
When $\bB(r_{12})$ is definite positive, the unique invariant measure of~\eqref{eq:eff_dyn} is $\nu(dv) = Z_\nu^{-1} \rme^{-U(v)} \, dv$. Recall that the argument $v$ is here a $d$-dimensional velocity, in constrast with~\eqref{eq:inv_meas_reduced_dyn_v} where only the parallel component $P^\pr(\br) v$ of the velocity was considered. 

\paragraph{Numerical integration} The numerical integration of~\eqref{eq:DPDE_elementary_anisotropic} for the choice~\eqref{eq:general_friction_fluctuation} can be performed as in Section~\ref{sec:Metropolis_elementary_FD}, except that matrix exponentials should be considered. Since the projection matrices are orthogonal and such that $P^\pr P^\perp = P^\perp P^\pr = 0$, the formulas for the proposed relative velocity $v_{12}^{n+1}$ simplify as $\widetilde{v}_{12}^{n+1} = P^\pr(\br_{12}) \widetilde{v}_{12}^{n+1} + P^\perp(\br_{12}) \widetilde{v}_{12}^{n+1}$ with
\[
\begin{aligned}
P^\pr(\br_{12}) \widetilde{v}_{12}^{n+1} & = \alpha^{n,\pr} P^\pr(\br_{12}) v_{12}^n + \eta^{n,\pr} \, P^\pr(\br_{12})\bG^n, \\
P^\perp(\br_{12}) \widetilde{v}_{12}^{n+1} & = \alpha^{n,\perp} P^\perp(\br_{12}) v_{12}^n + \eta^{n,\perp} \, P^\perp(\br_{12})\bG^n, \\
\end{aligned}
\]
where $\bG^n$ is a sequence of independent and identically distributed standard $d$-dimensional Gaussian random variables, and 
\begin{equation}
\label{eq:eff_coeff_step_n_general_FD}
\begin{aligned}
\alpha^{n,\pr} & = \exp\left(-\gamma^{\pr,n} \frac{\chi^\pr(r_{12})^2}{\mu_{12}}\dt \right), 
\qquad 
\eta^{n,\pr} = \sigma^\pr \sqrt{\frac{1-\left(\alpha^{n,\pr}\right)^2}{2 \gamma^{\pr,n} \mu_{12}}}, \\
\alpha^{n,\perp} & = \exp\left(-\gamma^{\perp,n} \frac{\chi^\perp(r_{12})^2}{\mu_{12}}\dt \right), 
\qquad 
\eta^{n,\perp} = \sigma^\perp \sqrt{\frac{1-\left(\alpha^{n,\perp}\right)^2}{2 \gamma^{\perp,n} \mu_{12}}},
\end{aligned}
\end{equation}
with $\gamma^{n,\pr} = \gamma^\pr(\eps_1^n,\eps_2^n)$ and $\gamma^{n,\perp} = \gamma^\perp(\eps_1^n,\eps_2^n)$ the friction coefficients at iteration~$n$. The probability of obtaining a new velocity $v'$ starting from $v^n$, which generalizes~\eqref{eq:transition_kernel_dt}, is therefore
\[
T_\dt(v^n,v') = \frac{1}{(2\pi)^{d/2}\eta^{n,\pr} (\eta^{n,\perp})^{d-1}} \exp\left(-\frac{\left|P^\pr(\br_{12})\left(v'-\alpha^{n,\pr} v^n\right)\right|^2}{2(\eta^{n,\pr})^2}-\frac{\left|P^\perp(\br_{12})\left(v'-\alpha^{n,\perp} v^n\right)\right|^2}{2(\eta^{n,\perp})^2}\right).
\]
When the proposed new energies 
\[
\widetilde{\eps}_1^{n+1} = \eps_1^n - \frac{\mu_{12}}{4}\left[\left(\widetilde{v}^{n+1}\right)^2 - \left(v_{12}^n\right)\right],
\qquad
\widetilde{\eps}_2^{n+1} = \eps_2^n - \frac{\mu_{12}}{4}\left[\left(\widetilde{v}^{n+1}\right)^2 - \left(v_{12}^n\right)\right]
\]
are positive, the logarithmic acceptance ratio in~\eqref{eq:expression_A_dt} can be computed. It reads
\[
\begin{aligned}
a_\dt(v^n,\widetilde{v}^{n+1}) & = s_1\!\left(\widetilde{\eps}_1^{n+1}\right) + s_2\!\left(\widetilde{\eps}_2^{n+1}\right)- s_1(\eps_1^n) - s_2(\eps_2^n) \\
& \ \ + \frac{|\bG^n|^2}{2} + \log \eta^{n,\pr} + (d-1)\log \eta^{n,\perp} - \log \widetilde{\eta}^{n+1,\pr} - (d-1)\log \widetilde{\eta}^{n+1,\perp}  \\
& \ \ -\frac{\left|P^\pr(\br_{12})\left(v^n-\widetilde{\alpha}^{n+1,\pr} \widetilde{v}^{n+1}\right)\right|^2}{2(\widetilde{\eta}^{n+1,\pr})^2}-\frac{\left|P^\perp(\br_{12})\left(v^n-\widetilde{\alpha}^{n+1,\perp} \widetilde{v}^{n+1}\right)\right|^2}{2(\widetilde{\eta}^{n+1,\perp})^2},
\end{aligned} 
\]
where $\widetilde{\alpha}^{n+1,\pr},\widetilde{\alpha}^{n+1,\perp},\widetilde{\eta}^{n+1,\pr},\widetilde{\eta}^{n+1,\perp}$ are defined as in~\eqref{eq:eff_coeff_step_n_general_FD} but with frictions evaluated at the proposed energies $\widetilde{\eps}_1^{n+1},\widetilde{\eps}_2^{n+1}$. Apart from these modifications, the algorithm summarized at the end of Section~\ref{sec:Metropolis_elementary_FD} is unchanged.


\end{document}